\documentclass[a4paper,usenatbib]{mnras2}

\usepackage{times}
\usepackage{graphicx}	% Including figure files
\usepackage{amsmath}	% Advanced maths commands
\usepackage{amssymb}	% Extra maths symbols
\usepackage{amsfonts}
\usepackage{color}

\usepackage{url}
\usepackage[flushleft]{threeparttable}

\usepackage{etoolbox}

\usepackage{placeins}

\hypersetup{colorlinks=true,linkcolor=blue,citecolor=blue,filecolor=blue,urlcolor=blue}
\makeatletter
\patchcmd{\NAT@citex}
  {\@citea\NAT@hyper@{%
     \NAT@nmfmt{\NAT@nm}%
     \hyper@natlinkbreak{\NAT@aysep\NAT@spacechar}{\@citeb\@extra@b@citeb}%
     \NAT@date}}
  {\@citea\NAT@nmfmt{\NAT@nm}%
   \NAT@aysep\NAT@spacechar\NAT@hyper@{\NAT@date}}{}{}

\patchcmd{\NAT@citex}
  {\@citea\NAT@hyper@{%
     \NAT@nmfmt{\NAT@nm}%
     \hyper@natlinkbreak{\NAT@spacechar\NAT@@open\if*#1*\else#1\NAT@spacechar\fi}%
       {\@citeb\@extra@b@citeb}%
     \NAT@date}}
  {\@citea\NAT@nmfmt{\NAT@nm}%
   \NAT@spacechar\NAT@@open\if*#1*\else#1\NAT@spacechar\fi\NAT@hyper@{\NAT@date}}
  {}{}
  
\usepackage{xspace}
  
\def\aj{AJ}
\def\araa{ARA\&A}
\def\apj{ApJ}
\def\apjl{ApJ}
\def\apjs{ApJS}
\def\ao{Appl.~Opt.}
\def\apss{Ap\&SS}
\def\aap{A\&A}

\def\aaps{A\&AS}

\def\mnras{MNRAS}

\def\pasa{Publ.~Astron.~Soc.~Aust.}
\def\pasp{PASP}
\def\pasj{PASJ}

\def\physscr{Phys.~Scr}

\def\rmxaa{RMxAA}

\def\assp{Astrophys.~Space~Sci.~Proc.}
\def\adndt{Atom.~Data~Nucl.~Data~Tabl.}
\def\cajpj{Can.~J.~Phys.}

\def\arcsec{\hbox{$^{\prime\prime}$}}

\newcommand{\foiii}{[O\,{\sc iii}]}

\newcommand{\foi}{[O\,{\sc i}]}
\newcommand{\foii}{[O\,{\sc ii}]}

\newcommand{\fsii}{[S\,{\sc ii}]}

\newcommand{\fnii}{[N\,{\sc ii}]}

\newcommand{\fariii}{[Ar\,{\sc iii}]}
\newcommand{\fariv}{[Ar\,{\sc iv}]}

\newcommand{\fcliii}{[Cl\,{\sc iii}]}
\newcommand{\fclii}{[Cl\,{\sc ii}]}

\newcommand{\nii}{N\,{\sc ii}}

\newcommand{\oii}{O\,{\sc ii}}

\newcommand{\ariv}{Ar\,{\sc iv}}

\newcommand{\hei}{He\,{\sc i}}
\newcommand{\heii}{He\,{\sc ii}}

\newcommand{\ha}{H$\alpha$}
\newcommand{\hb}{H$\beta$}
\newcommand{\hg}{H$\gamma$}

\newcommand{\bc}[1]{\textrm{\footnotesize{\color{blue}#1}}}

\newcommand{\orcidicon}{\includegraphics[width=0.26cm]{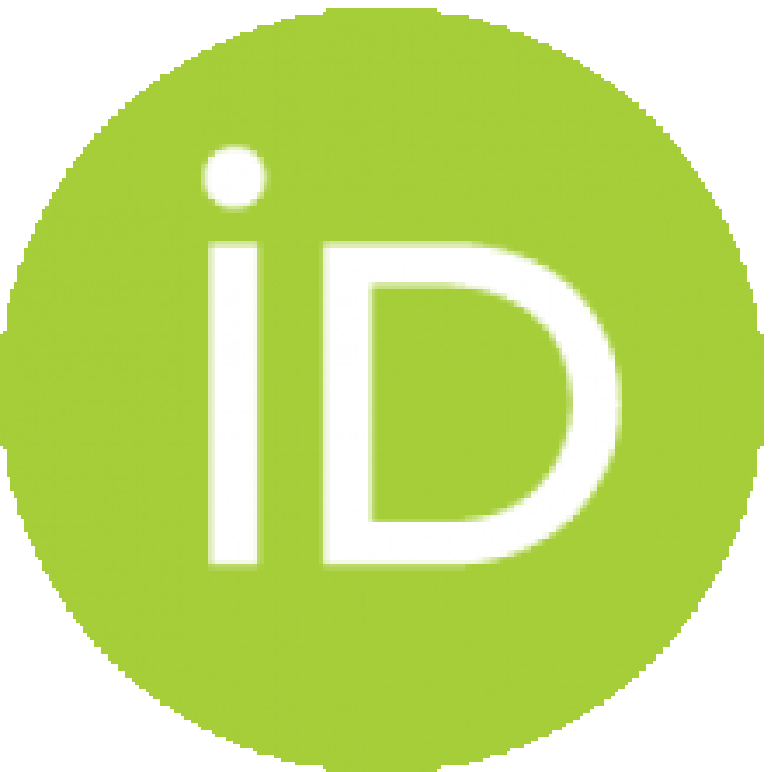}}

\newcommand{\orcidauthor}[1]{\,$^{\href{https://orcid.org/#1}{\orcidicon}}$}

\title[Chemical Abundances of IC\,4997]{Physical Conditions and Chemical Abundances of the Variable Planetary Nebula IC 4997}

\author[A.~Danehkar and M. Parthasarathy]{A.~Danehkar\orcidauthor{0000-0003-4552-5997}$^{{\bc{1},\bc{2}}}$\thanks{E-mail: \href{mailto:danehkar@eurekasci.com}{danehkar@eurekasci.com}} and
M. Parthasarathy\orcidauthor{0000-0002-8361-431X}$^{{\bc{3}}}$
\\
$^{1}$Department of Astronomy, University of Michigan, 1085 S. University Ave., Ann Arbor, MI 48109, USA\\
$^{2}$Eureka Scientific, 2452 Delmer Street Suite 100, Oakland, CA 94602-3017, USA\\
$^{3}$Indian Institute of Astrophysics, Koramangala, II Block, Bangalore, 560 034, India\\
}

\date{Accepted 2022 May 12. Received 2022 May 11; in original form 2021 August 17}

\pubyear{2022}

\begin{document}
\label{firstpage}

\pagerange{\pageref{firstpage}--\pageref{lastpage}} \pubyear{2022}

\maketitle

% Background: background information (present tense)
% Aim: principal activity (past/present perfect tense)
% Method: methodology (past tense)
% Results: results (past tense)
% Conclusion: conclusions (present tense/tentative verbs/model auxiliaries)
\begin{abstract}
The planetary nebula (PN) IC 4997 is one of a few rapidly evolving objects with variable brightness and nebular emission around a hydrogen-deficient star. In this study, we have determined the physical conditions and chemical abundances of this object using the collisionally excited lines (CELs) and optical recombination lines (ORLs) measured from the medium-resolution spectra taken in July 2014 with the FIbre-fed \'{E}chelle Spectrograph on the Nordic Optical Telescope at La Palma Observatory. We derived electron densities of $\gtrsim 3 \times 10^4$\,cm$^{-3}$ and electron temperatures  of $\gtrsim 14,000$\,K from CELs, whereas cooler temperatures of $\sim 11,000$ and $\sim 7,000$\,K were obtained from helium and heavy element ORLs, respectively. The elemental abundances deduced from CELs point to a metal-poor progenitor with [O/H]~$\lesssim$~$-0.75$, whereas the ORL abundances are slightly above the solar metallicity, [O/H]~$\approx$~$0.15$. Our abundance analysis indicates that the abundance discrepancy factors (ADFs\,$\equiv$\,ORLs/CELs) of this PN are relatively large: ADF(O$^{2+})\gtrsim 8$ and ADF(N$^{2+})\gtrsim 7$. Further research is needed to find out how the ADFs and variable emissions are formed in this object and whether they are associated with a binary companion or a very late thermal pulse.
\end{abstract}

% Select between one and six entries from the list of approved keywords.
% Don't make up new ones.

%%%%%%%%%%%%%%%%%%%%%%%%%%%%%%%%%%%%%%%%%%%%%%%%%%

%%%%%%%%%%%%%%%%% BODY OF PAPER %%%%%%%%%%%%%%%%%%

\section{Introduction}
\label{ic4997:introduction}

Asymptotic giant branch (AGB) stars with intermediate progenitor masses (1--$8M_{\odot}$) experience multiple thermal pulses at the end of their nuclear-burning stage, resulting in the formation of planetary nebulae (PNe). These thermal pulses usually happen during the AGB phase, when the helium-burning stellar shell is heated up \citep{Iben1975,Iben1983,Iben1983b}. This causes hydrogen-rich material to be pushed into the interstellar medium (ISM).

IC\,4997 (= PNG 058.3$-$10.9 = PK 58$-$10.1 = Hen~2-464 = VV~256 = IRAS~20178+1634) was discovered as a PN by \citet{Pickering1896}. The radio continuum observations of IC 4997 hinted at a double-shell morphology consisting of a compact inner shell of $0.3$ arcsec, an hourglass outer shell of $2.7 \times 1.4$ arcsec$^{2}$ with a position angle (PA) of 54$^{\circ}$, and bipolar jet-like components \citep{Miranda1996,Miranda1998}. Additionally, \citet{Soker2002} described it as a bent bipolar PN, which could be linked to a wide binary companion or a triple stellar system \citep{Bear2017}.

This PN has an unusually wide electron density range of $10^4$--$10^7$\,cm$^{-3}$ \citep{Aller1966,Aller1976,Ferland1979,Pottasch1984a,Hyung1994,Kostyakova1999,Zhang2004,Kostyakova2009,Burlak2010}. 
The derived electron density of the inner shell is very high in the range of $1.4 \times 10^5$--$2.6 \times 10^6$ cm$^{-3}$, while lower values of $1.0 \times 10^4$--$1.2 \times 10^4$ cm$^{-3}$ are found in the outer shell \citep{Miranda1996,Gomez2002}. 
Furthermore, electron temperatures have been reported in the $9.6 \times 10^3$ to $3.2 \times 10^4$\,K range  \citep{Menzel1941,Aller1966,Ferland1979,Flower1980,Pottasch1984a,Hyung1994,Kostyakova1999,Zhang2004,Kostyakova2009}. 

\begin{table*}
%\centering
\caption{NOT/FIES observation log of IC~4997 (Program ID: DDT2014-044, PI: J.\,J. D{\'{\i}}az-Luis).\label{tab:fies:obs:log}}
\begin{center}
\begin{tabular}{lccccll}
\noalign{\smallskip}
  \hline\noalign{\smallskip}
Seq. & Obs. Date \& Time  & Exp. (sec)  & RA  & DEC  &  Air mass  &  Target  \\
\noalign{\smallskip}
   \hline\noalign{\smallskip}
060072 &  2014-07-06 22:50:54 & 1800 & 20 20 08.74 & +16 43 53.69 & 1.62 & IC 4997 \\
\noalign{\smallskip}
060073 &  2014-07-06 23:22:21 & 1800 & 20 20 08.74 & +16 43 53.69 & 1.42 & IC 4997 \\
\noalign{\smallskip}
060074 &  2014-07-06 23:53:47 & 1800 & 20 20 08.74 & +16 43 53.69 & 1.27 & IC 4997 \\
\noalign{\smallskip}
060075 &  2014-07-07 00:25:14 & 1800 & 20 20 08.74 & +16 43 53.69 & 1.17 & IC 4997 \\
\noalign{\smallskip}
060076 &  2014-07-07 00:56:40 & 1800 & 20 20 08.74 & +16 43 53.69 & 1.10 & IC 4997 \\
\noalign{\smallskip}
060077 &  2014-07-07 01:28:07 & 1800 & 20 20 08.74 & +16 43 53.69 & 1.06 & IC 4997 \\
\noalign{\smallskip}
060078 &  2014-07-07 01:59:33 & 1800 & 20 20 08.74 & +16 43 53.69 & 1.03 & IC 4997 \\
\noalign{\smallskip}
060079 &  2014-07-07 02:31:01 & 1800 & 20 20 08.74 & +16 43 53.69 & 1.02 & IC 4997 \\
\noalign{\smallskip}
060080 &  2014-07-07 03:05:59  & 180 & 20 20 08.74 & +16 43 53.69 & 1.03 & IC 4997 \\
\noalign{\smallskip}
060064 &  2014-07-06 21:28:20 & 300  & 15 50 02.89 & +33 05 48.86 & 1.01 & BD+33$^{\circ}$2642 \\
\noalign{\smallskip}\hline
\end{tabular}
\end{center}
\end{table*}

The central star of IC\,4997 has been classified among hydrogen-deficient stars: [WC7/8] \citep{Swings1941}, Of-WR \citep{Smith1969,Aller1975a},  weak-emission line star \citep{Tylenda1993,Marcolino2003,Marcolino2007}, and [WC]-PG1159 \citep{Parthasarathy1998}. Nevertheless, a recent study of the stellar spectrum of IC\,4997 with Gemini Spectroscopy in 2014 did not lead to any classification owing to the presence of photospheric H line and nebular contamination \citep{Weidmann2015}. 

The origins of hydrogen-deficient central stars of planetary nebulae (CSPNe) are not well understood. A (very-) late thermal pulse (LTP/VLTP) during the post-AGB and early PN stages, known as the born-again scenario, has been proposed to lead to the formation of a hydrogen-deficient star \citep{Bloecker2001,Herwig2001,Koesterke2001,Werner2001,Werner2006}. A born-again event could also eject metal-rich, dense, small-scale structures into the previously expelled hydrogen-rich gas. It is well known that abundances derived from optical recombination lines (ORLs) are usually higher than those from collisionally excited lines (CELs) in several PNe \citep[e.g.][]{Liu2000,Liu2001,Liu2004a,Wesson2003,Tsamis2003,Tsamis2004,Tsamis2008,Garcia-Rojas2009,Garcia-Rojas2013,Danehkar2021}, while temperatures measured from ORLs are typically lower than those from CELs  \citep[e.g.][]{Peimbert1967,Peimbert1971,Wesson2004,Wesson2005}. Tiny metal-rich, dense knots embedded in the hydrogen-rich PN may be to blame for the difference in ORL and CEL abundances \citep{Liu2003,Liu2004b}. 

Variable nebula emissions \citep{Aller1966,Ferland1979,Feibelman1979,Feibelman1992,Kostyakova1999,Kostyakova2009,Burlak2010,Arkhipova2020}, photometric variability of the nebula brightness \citep{Arkhipova1994,Kostyakova2009,Arkhipova2020}, and changes in the effective temperature of the central star \citep{Feibelman1979,Purgathofer1981,Kiser1982,Kostyakova1999,Kostyakova2009} make IC 4997 a unique object. A decline in the effective temperature of the central star from $T_{\rm eff}=59$\,000 K to 47\,000\,K has been found over the period 1895--1962 \citep{Feibelman1979}. Moreover, there are changes in the flux ratio \foiii\,$\lambda$4363/\hg\,$\lambda$4340, which were interpreted as a gradual increase in the stellar temperature from 1962 until 1977 and an outburst (or flare) in 1977--78 \citep{Feibelman1979,Purgathofer1981}. However, \citet{Kiser1982} suggested that the changes in nebula emission lines could be related to the expansion of the nebula itself. Furthermore, an increase in the stellar temperature from $T_{\rm eff}\approx40$\,000 to 60\,000\,K  over the period 1972--1992 was proposed to be responsible for the variability in the nebular spectrum \citep{Kostyakova1999}. Nevertheless, the average temperature of the central star, which was estimated from the nebular continuum, corresponds to $T_{\rm eff}=37$\,000\,K in 1972--1974 and 47\,500\,K in 1999--2005 \citep{Kostyakova2009}, so the stellar temperature has likely increased by about 10,000 K in three decades. More recently, \citet{Miranda2022} proposed that the main reason for the variability in this object could be an episodic, smoothly changing stellar wind based on a relationship between the stellar wind strength and the \foiii\ $\lambda$4363/H$\gamma$ line ratio.

In this paper, we present our study of the spectroscopic measurements of the PN~IC\,4997 taken in 2014 with the 2.56-m Nordic Optical Telescope, with the aim of obtaining physical conditions and chemical abundances from CELs and ORLs. We also use the \textit{International Ultraviolet Explorer} (\textit{IUE}) spectra to characterize the ionizing UV luminosity of the central star. This paper is organized as follows: Section \ref{ic4997:observations} describes the details of the observations. Section \ref{ic4997:results:flux} explains the emission line identification and flux measurements. In Sections \ref{ic4997:temp:dens} and \ref{ic4997:abund}, we present the results from our plasma diagnostics and abundance analysis, respectively, followed by a discussion of stellar characteristics in Section \ref{ic4997:stellar}. In Section \ref{ic4997:variability}, we discuss the variability of emission lines in this PN. Finally, our findings are summarized in Section \ref{ic4997:summary}. 

\begin{figure*}
\begin{center}
\includegraphics[width=6.5in]{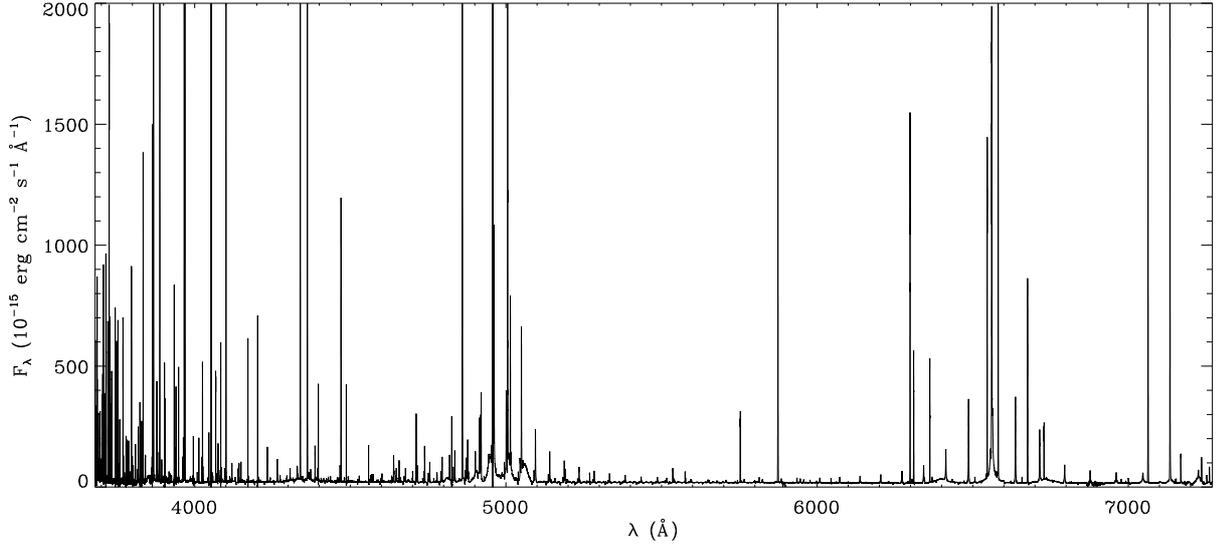}
\end{center}
\caption{The NOT/FIES spectrum of IC\,4997 collected with the exposure time of 1800 seconds in the wavelength range of 3680--7270 {\AA}.
\label{fig:ic4997:spectrum}}
\end{figure*}

\begin{table*}
\caption{\textit{International Ultraviolet Explorer} Observations of IC~4997 (low-dispersion spectra). \label{tab:iue:obs:log}}
\begin{center}
\begin{tabular}{lllllcl}
\noalign{\smallskip}
  \hline\noalign{\smallskip}
Instrument	& Obs.~ID	&Program ID \& PI	&Obs.~Date	& Exposure (s)	& $\lambda($\AA$)$ \\
\noalign{\smallskip}
\hline
\noalign{\smallskip}
SWP	&swp01732	&PN2AB, Boggess	&1978 June 7	&1800	&1150--1975\\
\noalign{\smallskip}
LWR	&lwr01628	&PN2AB, Boggess	&1978 June 7	&1200	&1975--3300\\
\noalign{\smallskip}
SWP	&swp08578	&NPCAB, Boggess	&1980 March 28	&2400	&1150--1975\\
\noalign{\smallskip}
LWR	&lwr07317	&NPCAB, Boggess	&1980 March 28	&1800	&1975--3300\\
\noalign{\smallskip}
SWP	&swp14400	&NPDWF, Feibelman	&1981 July 5	&1800	&1150--1975\\
\noalign{\smallskip}
LWR	&lwr11011	&NPDWF, Feibelman	&1981 July 5	&1800	&1975--3300\\
\noalign{\smallskip}
SWP	&swp31683	&NPJGF, Ferland	&1987 Sept. 1	&1800	&1150--1975\\
\noalign{\smallskip}
LWP	&lwp11540	&NPJGF, Ferland	&1987 Sept. 1	&1800	&1975--3300\\
\noalign{\smallskip}
SWP	&swp41945	&PNNLA, Aller	&1991 June 29	&1500	&1150--1975\\
\noalign{\smallskip}
LWP	&lwp20705	&PNNLA, Aller	&1991 June 29	&1800	&1975--3300\\
\noalign{\smallskip}\hline
\end{tabular}
\end{center}
\end{table*}

\section{Observations}
\label{ic4997:observations}

The PN IC\,4997 was observed on July 7th, 2014 (PI: J.\,J. D{\'{\i}}az-Luis; Program ID: DDT2014-044) using the FIbre-fed \'{E}chelle Spectrograph \citep[FIES;][]{Telting2014} on the 2.56-m Nordic Optical Telescope \citep[NOT;][]{Djupvik2010} at the Roque de los Muchachos Observatory in La Palma, Spain. The observing run was a part of a program detecting diffuse interstellar bands and fullerene-based molecules in PNe \citep{Diaz-Luis2016}. The observation log is presented in Table~\ref{tab:fies:obs:log}. The observations were made using the bundle C (F3) medium-resolution mode ($R = 46\,000$) with the 1.3$\arcsec$ fibre located on the PN IC\,4997, covering the wavelength range 3640--7270\,{\AA}. Exposure times of 180 and 1800 seconds were used, which yielded signal-to-noise ratios of at least 30 towards the blue-end and more than 60 towards the red-end of the spectrum. We combined all 8 observations taken with an exposure time of 1800 seconds (see Table~\ref{tab:fies:obs:log}) using the {\sc iraf} task {\sf scombine}. Most of the bright lines were saturated in the combined 1800-sec spectrum, so we employed it for the measurements of faint lines. We utilized the 180-sec spectrum to measure the bright lines. The \ha\ $\lambda$6563 and \foiii\ $\lambda$5007 emission lines were saturated in the exposure of 180 sec, so they could not be used for our study. The FIEStool\footnote{\href{http://www.not.iac.es/instruments/fies/fiestool/}{http://www.not.iac.es/instruments/fies/fiestool/}} Python package was employed to reduce the data. The data reduction steps include bias subtraction, scattering subtraction, flat-field correction, one-dimensional spectrum extraction, blaze-shape correction, and wavelength calibration. The spectrophotometric standard star BD+33$^{\circ}$2642 (SP1550+330) was used to perform the flux calibration using the {\sc iraf} standard functions. The averaged NOT/FIES spectrum of this PN is presented in Figure \ref{fig:ic4997:spectrum}, which shows several ORLs.

\begin{table}
\caption{The extinction $c({\rm H}\beta)$ derived from the Balmer lines. \label{tab:result:ext}}
\begin{center}
\begin{tabular}{lccl}
\noalign{\smallskip}
\hline
\noalign{\smallskip}
Line & Mult & Weight & $c({\rm H}\beta)$ \\
\noalign{\smallskip}
\hline
\noalign{\smallskip}
$\lambda$3697.15  & B17  &    1 & $ 0.552_{-0.015}^{+ 0.013}$ \\
\noalign{\smallskip}
$\lambda$3711.97  & B15  &    1 & $ 0.722_{-0.056}^{+ 0.050}$ \\
\noalign{\smallskip}
$\lambda$3734.37  & B13  &    2 & $ 0.586_{-0.012}^{+ 0.010}$ \\
\noalign{\smallskip}
$\lambda$3750.15  & B12  &    3 & $ 0.480_{-0.020}^{+ 0.017}$ \\
\noalign{\smallskip}
$\lambda$3770.63  & B11  &    3 & $ 0.710_{-0.010}^{+ 0.008}$ \\
\noalign{\smallskip}
$\lambda$3797.90  & B10  &    5 & $ 0.747_{-0.010}^{+ 0.008}$ \\
\noalign{\smallskip}
$\lambda$3835.39  & B9   &    6 & $ 0.783_{-0.010}^{+ 0.008}$ \\
\noalign{\smallskip}
$\lambda$3970.07  & B7   &   14 & $ 0.534_{-0.007}^{+ 0.006}$ \\
\noalign{\smallskip}
$\lambda$4101.74  & B6   &   22 & $ 0.622_{-0.022}^{+ 0.018}$ \\
\noalign{\smallskip}
$\lambda$4340.47  & B5   &   40 & $ 0.674_{-0.040}^{+ 0.034}$ \\
\noalign{\smallskip}
\hline
\noalign{\smallskip}
Average             &  &  &  $ 0.645_{-0.022}^{+ 0.018}$  \\
\noalign{\smallskip}
\noalign{\smallskip}
\hline
\end{tabular}
\end{center}
\small
\begin{tablenotes}
\item[1]\textbf{Note.} The weights calculated based on the \hei\ theoretical predictions are used for calculating the average extinction. 
The uncertainty of the average extinction obtained from weighted-uncertainties using an MCMC-based method at a confidence level of $90\%$. 
\end{tablenotes}
\end{table}

The \textit{IUE} UV spectra of IC\,4997 were retrieved from the Mikulski Archive for Space Telescopes (MAST). An observational journal of the \textit{IUE} data is provided in Table~\ref{tab:iue:obs:log}. As we aimed to estimate the UV luminosities from spectral continua, we searched for low-dispersion spectroscopic observations of this object and used the highest exposure spectrum when there were multiple exposures in each observing program. The \textit{IUE} spectra were taken by the short-wavelength prime (SWP) and the long-wavelength redundant (LWR) cameras, which cover the wavelength ranges of 1150--1975\,\AA\ and 1975--3300\,\AA, respectively. The \textit{IUE} observed IC\,4997 in different epochs during 13 years, namely 1978, 1980, 1981, 1987, and 1991, when this object had very high variabilities in the \hei\ $\lambda$4471 line flux and the \foiii\ $\lambda$4363/$\lambda$4959 flux ratio due to possible outbursts (see \S\,\ref{ic4997:variability}). The \foiii\ flux measurements indicate that the stellar temperature was decreasing after 1973 until it reached its lowest value in 1986, while it was gradually increasing in the following years, so the \textit{IUE} observations cover this period of the remarkable shift in the stellar temperature. 

\section{Emission Line Measurements}
\label{ic4997:results:flux}

We automatically identified and measured emission lines with the IDL package MGFIT,\footnote{\href{https://github.com/mgfit/MGFIT-idl}{https://github.com/mgfit/MGFIT-idl}} which applies the Levenberg-Marquardt least-squares minimization \citep{Markwardt2009} to multiple Gaussian functions, and optimizes the solutions using a genetic algorithm \citep{Wesson2016} and a random walk technique. We manually verified and removed any misidentified lines. To estimate the errors of emission lines, the signal-dependent noise model of the Gaussian fit \citep{Landman1982,Lenz1992} was used to quantify the RMS deviation of the continuum around each emission line. 
Our bright line fluxes with respect to ${\rm H}\beta$ measured from the NOT observations are roughly similar to the fluxes in 2013 and 2015 observed by \citet{Arkhipova2020}, though they might observe a different region of the nebula.

We obtained the dust and ISM extinction at H$\beta$ from the 10 observed Balmer emission lines relative to ${\rm H}\beta$ listed in Table~\ref{tab:result:ext} using their theoretical predictions under the Case B assumption \citep{Storey1995} and the Galactic reddening functions of  \citet[][$\lambda <5465$\,{\AA}]{Seaton1979a} and \citet[][$\lambda >5465$\,{\AA}]{Howarth1983}. The physical conditions for the extinction calculations were numerically obtained by iterations seeking the convergence of $N_{\rm e}([$O~{\sc ii}]) and $T_{\rm e}([$N~{\sc ii}]) according to \cite{Ueta2021}'s prescriptions. 
The flux measurement errors in the Balmer emission lines were also propagated into the extinction analysis without taking into account the physical condition uncertainties. The logarithmic extinctions $c({\rm H}\beta)$ estimated from different Balmer emission lines were weighted based on their predicted intrinsic intensities relative to ${\rm H}\beta$ for the derived physical condition. As a result, $c({\rm H}\beta)$ derived from weak Balmer lines has lower weights, whereas $c({\rm H}\beta)$ derived from strong Balmer lines has higher weights. The \ha\ $\lambda$6563 line was saturated, so we could not use it for the extinction calculation. 

To deredden the observed fluxes of the emission lines, we used the weighted average $c({\rm H}\beta)= 0.645$. Our averaged extinction agrees with previous values of $c({\rm H}\beta) = 0.61$ \citep{Odell1963}, whereas $0.8$ \citep{Hyung1994}, $0.32$ \citep{Ruiz-Escobedo2022}, and other values were also reported \citep[][]{Ahern1978,Flower1980,Zhang2004,Kostyakova2009,Burlak2010,Arkhipova2020}. Taking $F({\rm H}\alpha)/F({\rm H}\beta)= 4.25$ as measured in 2015 by \citet{Arkhipova2020}, we get $c({\rm H}\beta) =0.59$ for the theoretical ${\rm H}\alpha/{\rm H}\beta$ prediction with our physical conditions $N_{\rm e}([$O~{\sc ii}]) and $T_{\rm e}([$N~{\sc ii}]). The differences in the extinction values measured by different observations can be explained by the inhomogeneous distribution of the dust extinction over the nebula, as seen in the radio 3.6-cm continuum map taken in 1995 \citep[][]{Miranda1996}. Spatial variations of the extinction $c({\rm H}\beta)$ have been found in other PNe \citep[see e.g.][]{Monteiro2004,Monteiro2005,Lee2005,Akras2020,Otsuka2022}. Moreover, the Galactic dust reddening and extinction service\footnote{\href{https://irsa.ipac.caltech.edu/applications/DUST/}{https://irsa.ipac.caltech.edu/applications/DUST/}}
yields $E(B-V)= 0.136 \pm 0.006$ \citep{Schlegel1998} and $0.117 \pm 0.005$ \citep{Schlafly2011} with a 10-arcmin circular aperture centered on IC\,4997, which could correspond to the line-of-sight Galactic extinction.

\begin{table*}
\caption{References for atomic data.
\label{tab:atomicdata}
}
\footnotesize
\centering
\begin{tabular}{lll}
\hline
\noalign{\smallskip}
Ion 	& Transition probabilities & Collision strengths \\
\noalign{\smallskip}
\hline 
\noalign{\smallskip} 
N${}^{+}$   & \citet{Bell1995} & \citet{Stafford1994} \\ 
\noalign{\smallskip} 
O${}^{0}$   & \citet{FroeseFischer2004} & \citet{Zatsarinny2003}, \citet{Bell1998} \\ 
O${}^{+}$   & \citet{Landi2012} & \citet{Pradhan2006}, \citet{McLaughlin1994} \\ 
O${}^{2+}$  & \citet{Storey2000}, \citet{Tachiev2001} &  \citet{Lennon1994}, \citet{Bhatia1993} \\ 
\noalign{\smallskip} 
Ne${}^{2+}$ & \citet{Daw2000}, \citet{Landi2005}  &  \citet{McLaughlin2000}\\ 
\noalign{\smallskip} 
S${}^{+}$   & Nahar (unpublished, 2001) & \citet{Ramsbottom1996}  \\ 
S${}^{2+}$  & \citet{Tayal1997}, \cite{Huang1985} & \citet{Tayal1997}, \citet{Tayal1999} \\
\noalign{\smallskip} 
Cl${}^{2+}$  & \citet{Mendoza1982} &  \cite{Ramsbottom2001} \\ 
\noalign{\smallskip} 
Ar${}^{2+}$ & \citet{Biemont1986} & \citet{Galavis1995} \\ 
Ar${}^{3+}$ & \citet{Landi2012} & \citet{Ramsbottom1997}, \citet{Ramsbottom1997a} \\ 
%Ar${}^{4+}$ & \citet{Biemont1983} & \citet{Galavis1995} \\ 
\noalign{\smallskip}  
Fe${}^{2+}$ & \citet{Ercolano2008} & \citet{Ercolano2008} \\ 
\noalign{\smallskip}
\hline
\noalign{\smallskip}
Ion 	& Effective recombination coefficients & Case \\
\noalign{\smallskip}
\hline 
\noalign{\smallskip} 
H${}^{+}$   & \citet{Storey1995} &  B \\ 
\noalign{\smallskip} 
He${}^{+}$   & \citet{Porter2013} &  B \\ 
\noalign{\smallskip} 
He${}^{2+}$  & \citet{Storey1995} & B \\ 
\noalign{\smallskip} 
C${}^{2+}$  & \citet{Davey2000} & A,\,B \\ 
C${}^{3+}$  & \citet{Pequignot1991} & A \\ 
\noalign{\smallskip} 
N${}^{2+}$  & \citet{Fang2011,Fang2013a} & B \\ 
N${}^{3+}$  & \citet{Pequignot1991} & A \\ 
\noalign{\smallskip} 
O${}^{2+}$  & \citet{Storey2017} & B  \\ 
\noalign{\smallskip}  
\hline
\end{tabular}
\end{table*}

Dereddened emission line fluxes were calculated using the formula $I(\lambda)=10^{c({\rm H}\beta)[1+f(\lambda)]}\,F(\lambda)$, where $F(\lambda)$ is the observed flux, $I(\lambda)$ is the dereddened flux, and $f(\lambda)$ is the extinction function of the Galactic reddening law  \citep{Seaton1979a,Howarth1983} with $R_V\equiv A(V)/E(B-V)=3.1$, normalized so that $f(\lambda)=0$ at the {\rm H}$\beta$ wavelength ($\lambda=4861.33$\,{\AA}). In Table~A1 (Supplementary Material), we list the identified emission lines together with their measured fluxes and measurement errors. Columns 1--12 show the laboratory wavelength at ($\lambda_{\rm lab}$), the ion identification, the laboratory wavelength, the observed wavelength, the observed flux with measurement errors (in percentage), the dereddened flux with corresponding errors (in percentage), multiplet number (with the prefix `H' for hydrogen Balmer lines, `V' for ORLs, and `F' for CELs), the lower and upper terms of the transition, and the statistical weights of the lower and upper levels, respectively. All flux values are given with respect to the H$\beta$ flux, on a scale where ${\rm H}\beta=100$. 

We derived a systemic heliocentric velocity of $v_{\rm hel}=-69.2$\,km\,s$^{-1}$ from the \hb\ emission line in the NOT spectrum using the IDL task \textsc{rvcorrect}. This systemic velocity is in agreement with $v_{\rm hel}=-66.2$\,km\,s$^{-1}$ \citep{Durand1998} and $v_{\rm hel}=-64.4$\,km\,s$^{-1}$ \citep{Kharchenko2007}. Moreover, \citet{Frew2016} obtained a distance of $4.85 \pm 1.56$ kpc using the H$\alpha$ surface brightness-radius relation, whereas \citet{Bailer-Jones2018} estimated a distance of 4.32 kpc with a lower limit of 2.68 and an upper limit of 6.75 kpc. Considering the distance of $4.32_{-2.17}^{+1.90}$ kpc, the hourglass shell of $2.7 \times 1.4$\,arcsec$^{2}$ corresponds to the angular diameter of $0.057_{-0.019}^{+0.022} \times 0.029_{-0.010}^{+0.012}$\,parsec$^{2}$. The long-slit \fnii\ kinematic observation of IC 4997 in October 2003 \citep{Lopez2012b} also suggests a maximum expansion velocity of $60_{-8}^{+16}$ km\,s$^{-1}$ with respect to the central star, assuming an inclination of $i=35^{\circ}$ based on its hourglass morphology. Thus, the deprojected outflows could reach a distance of $0.035_{-0.012}^{+0.014}$\,parsec relative to the central star. The nebula size is comparable to Hen\,2-142 with a [WC\,9] star and Pe\,1-1 with a [WO\,4] star, but those PNe depict higher outflow velocities of $\sim 200$ km\,s$^{-1}$ \citep{Danehkar2022}. For the \fnii\ expansion velocity, we obtained a kinetic age of about $570_{-180}^{+200}$ years. However, the true age could be around 1.5 of the kinetic age \citep{Dopita1996}, so this PN could have a PN age of around $t_{\rm age} = 860_{-270}^{+300}$ years. 

\section{Physical Conditions}
\label{ic4997:temp:dens}

We carried out plasma diagnostics of the dereddened fluxes of CELs using the IDL library proEQUIB \citep{Danehkar2018}. 
To estimate errors, we employed an IDL implementation\footnote{\href{https://github.com/mcfit/idl_emcee}{https://github.com/mcfit/idl\_emcee}} (\textsf{idl\_emcee}) of the affine-invariant Markov chain Monte Carlo (MCMC) hammer \citep{Goodman2010}. %, which was also implemented in Python language \citep{Foreman-Mackey2013}. 
In the MCMC hammer, a confidence level of $90\%$ and a uniform uncertainty distribution were used to propagate flux errors into the plasma diagnostics and abundance analysis. For the CEL calculations, we mostly used the CHIANTI atomic data v7.0 \citep{Landi2012} read from the IDL library AtomNeb \citep{Danehkar2019}. The references for atomic data are listed in Table~\ref{tab:atomicdata}, which includes transition probabilities, collision strengths, and effective recombination coefficients. 

Table~\ref{tab:result:cel:temp:dens} lists the electron densities ($N_{\rm e}$) and electron temperatures ($T_{\rm e}$) derived from the available diagnostic CEL flux ratios. Following \cite{Ueta2021}, we performed the CEL plasma diagnostics iteratively to achieve the converged results for a pair of coupled $T_{\rm e}$- and $N_{\rm e}$-diagnostics. 
We linked $T_{\rm e}$(\fnii) to $N_{\rm e}$(\foii), and the \fariii\ temperature to $N_{\rm e}$(\fclii). 
As the \foiii\ temperature is higher than typical PN temperatures of $< 25,000$\,K at densities below $2 \times 10^5$\,cm$^{-3}$, we made a link between $T_{\rm e}$(\foiii) and $N_{\rm e}$(\ariv). The \foii\ density was also used to calculate the \foi\ and \fsii\ temperatures.

We also estimated the electron temperature $T_{\rm e,rc}$(\fnii) using the $\lambda$5755 auroral line corrected for the recombination contamination according to the empirical formula of  \citet{Liu2000}. To estimate the recombination contribution, we used the ionic abundance  ${\rm N^{2+}}$ derived from the N\,{\sc ii} ORLs in \S\,\ref{ic4997:abund}. The O$^{+3}$ ionic abundance estimated from ${\rm O^{3+}}/{\rm H^{+}}=\lbrack ({\rm He}/{\rm H^{+}})^{2/3} -1 \rbrack \times ( {\rm {\rm O^{+}}/{\rm H^{+}} + O^{2+}}/{\rm H^{+}} )$ is negligible, so it is not necessary to correct the [O\,{\sc iii}] $\lambda$4363 line flux for the recombination contamination.

\begin{figure*}
\begin{center}
\includegraphics[width=5.0in]{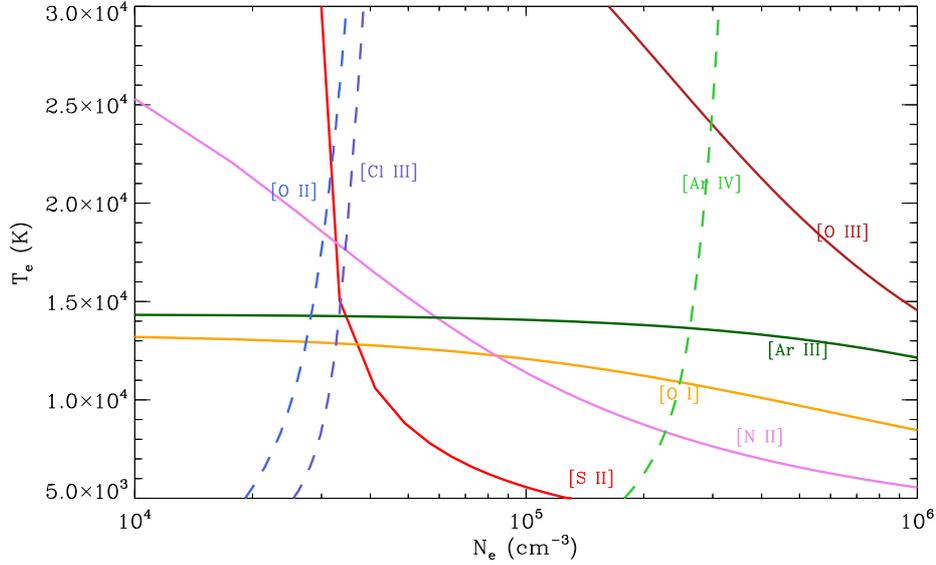}
\end{center}
\caption{$N_{\rm e}$--$T_{\rm e}$ diagnostic diagram for CELs.
The solid red lines correspond to $T_{\rm e}$-diagnostic curves, and the dashed blue lines correspond to $N_{\rm e}$-diagnostic curves with emission lines listed in Table~\ref{tab:result:cel:temp:dens}. The $T_{\rm e}$-diagnostic \fnii\ curve is without the recombination correction.
\label{ic4997:diagnostic:diagram}%
}%
\end{figure*}

Figure~\ref{ic4997:diagnostic:diagram} presents the $N_{\rm e}$--$T_{\rm e}$ diagnostic diagram made from the $N_{\rm e}$-diagnostic CEL ratios of the \foii, \fariv, and \fcliii\ ions, and the $T_{\rm e}$-diagnostic CEL ratios of the \foi, \fnii, \fsii, \foiii, and \fariii\ ions. The $T_{\rm e}$-diagnostic \fnii\ curve shown in this figure is based on the \fnii\ flux ratio, which was not corrected for the recombination contribution to its $\lambda$5755 auroral line. 
It can be seen that the electron temperatures derived from the \fnii, \fsii\ and \foiii\ flux ratios are very sensitive to the electron density, whereas the \foi\ and \fariii\ temperatures show less variability with $N_{\rm e}$. The \fariv\ $N_{\rm e}$-diagnostics also have a noticeable dependence on $T_{\rm e}$, whereas the electron densities derived from the \foii\ and \fcliii\ flux ratios are less sensitive to the electron temperature. 

It is important to consider the critical densities of diagnostic lines when we interpret the results. We calculated the critical densities for density-diagnostic low-excitation lines at $T_{\rm e} =$\,18,000\,K as follows: 
[S\,{\sc ii}] $\lambda\lambda$6717,\,6731, $N_{\rm cr}=$\,2,290, 6,770\,cm$^{-3}$; and 
[O\,{\sc ii}] $\lambda\lambda$3726,\,3729, $N_{\rm cr}=$\,4,930, 1,240\,cm$^{-3}$, respectively; 
whereas for high-excitation $N_{\rm e}$-diagnostic lines: 
[Ar\,{\sc iv}] $\lambda\lambda$4711,\,4740, $N_{\rm cr}=$\,20,480, 184,510\,cm$^{-3}$ ($T_{\rm e} =$\,24,100\,K); and  
[Cl\,{\sc iii}] $\lambda\lambda$5518,\,5538, $N_{\rm cr}=$\,6,820, 36,980\,cm$^{-3}$ ($T_{\rm e} =$\,14,300\,K), respectively. Although the \fariv\ density-diagnostic lines have the highest critical density, our derived \fariv\ density is highly uncertain, so the \fcliii\ density-diagnostic line ratio may be more reliable. Thus, we assumed the \fcliii\ density for the higher ionized ions (X$^{i+}$, $i \geqslant 2$) in our first analysis in Section~\ref{ic4997:abund}. To see the difference, we used the \fariv\ density for X$^{i+}$ ($i \geqslant 2$) in our second abundance analysis.

\begin{figure*}
\begin{center}
\includegraphics[width=2.8in]{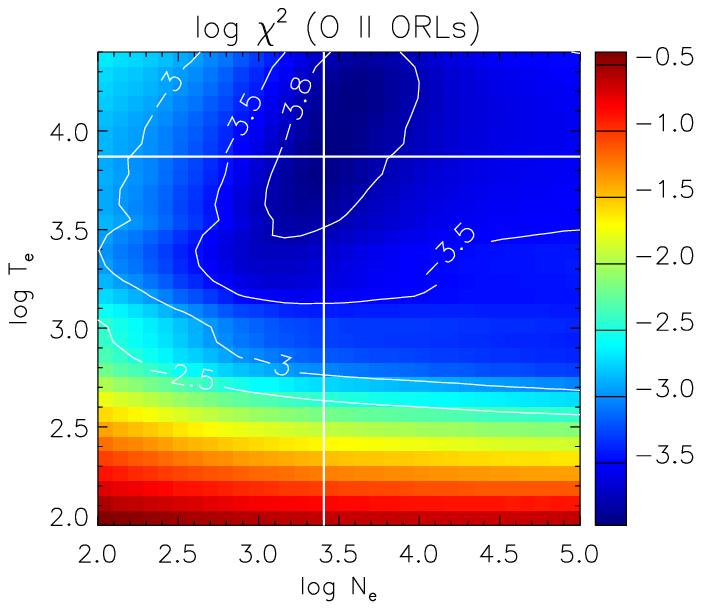}
\includegraphics[width=2.8in]{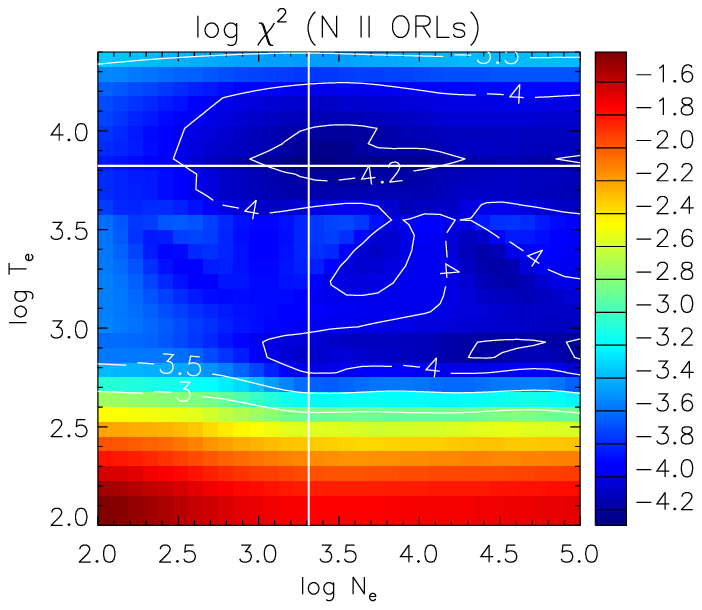}%
\end{center}
\caption{$N_{\rm e}$--$T_{\rm e}$ diagnostic diagrams based on O\,{\sc ii} and N\,{\sc ii} ORLs. %}
\label{fig:dd:orls}
}
\end{figure*}

\begin{table}
\caption{Physical conditions derived from CELs and ORLs. \label{tab:result:cel:temp:dens}}
\footnotesize
\begin{center}
\begin{tabular}{lcc}
\noalign{\smallskip}
\hline
\noalign{\smallskip}
 Lines  & Temperature Diagnostic & $T_{\rm e}$(K) \\
\noalign{\smallskip}
\hline
\noalign{\smallskip}
$[$O\,{\sc i}] CELs$^{\mathrm{a}}$     &$\frac{6300.34+6363.78}{5577.34}$              & $     12910_{      -240}^{+       300}$ \\
\noalign{\smallskip}
$[$N\,{\sc ii}] CELs$^{\mathrm{a}}$    &$\frac{6548.10+6583.50}{5754.60}$              & $     18480_{      -530}^{+       530}$ \\
\noalign{\smallskip}
$[$N\,{\sc ii}] CELs$^{\mathrm{a},\star}$    &$\frac{6548.10+6583.50}{5754.60}$             & $     17990_{      -490}^{+       520}$ \\
\noalign{\smallskip}
[S\,{\sc ii}]  CELs$^{\mathrm{a}}$ & $\frac{6716.44+6730.82}{4068.60+4076.35}$      & $     21260_{     -4560}^{}$  \\
\noalign{\smallskip}
$[$Ar\,{\sc iii}] CELs$^{\mathrm{b}}$  &$\frac{7135.80+7751.43}{5191.82}$              & $     14260_{      -290}^{+       270}$  \\
\noalign{\smallskip}
$[$O\,{\sc iii}] CELs$^{\mathrm{b}}$   &$\frac{4958.91}{4363.21}$              & $     24050_{      -370}^{+       470}$ \\
\noalign{\smallskip}
%\hline
%\noalign{\smallskip}
He\,{\sc i} ORLs      & $\frac{5875.66}{4471.50}$             & $     11070_{      -880}^{+      1010}$ \\
\noalign{\smallskip}
O\,{\sc ii} ORLs      & least-squares min.             & $      7400_{     -5300}^{+     17400}$ \\
\noalign{\smallskip}
N\,{\sc ii} ORLs      & least-squares min.             & $      6700_{     -2900}^{+      8100}$ \\
\noalign{\smallskip}
$\Delta T_{\rm +}{}^{\mathrm{c}}$      &  $T_{\textrm{[N\,{\sc ii}]}} - T_{\textrm{He\,{\sc i}}}$    &  $7410_{-1380}^{+1470} $ \\
\noalign{\smallskip}
$\Delta T_{\rm 2+}{}^{\mathrm{c}}$      &  $T_{\textrm{[Ar\,{\sc iii}]}} - T_{\langle \textrm{O\,{\sc ii}}, \textrm{N\,{\sc ii}} \rangle}$    &  $7210_{-7210}^{+4500} $ \\
\noalign{\smallskip}
\hline
\noalign{\smallskip}
 Lines  & Density Diagnostic & $N_{\rm e}$(cm$^{-3}$) \\
\noalign{\smallskip}
\hline
\noalign{\smallskip}
$[$O\,{\sc ii}] CELs$^{\mathrm{d}}$    &$\frac{3728.82}{3726.03}$             & $     30400_{    -13740}^{+     45550}$ \\
\noalign{\smallskip}
$[$Cl\,{\sc iii}] CELs$^{\mathrm{d}}$  &$\frac{5537.60}{5517.66}$              & $     33260_{     -2870}^{+      3300}$ \\
\noalign{\smallskip}
$[$Ar\,{\sc iv}] CELs$^{\mathrm{d}}$    &$\frac{4740.17}{4711.37}$              & $    297380_{    -34310}^{+     43730}$ \\
\noalign{\smallskip}
O\,{\sc ii} ORLs$^{\mathrm{e}}$       & least-squares min.             & $      2500_{     -1700}^{+     15200}$ \\
\noalign{\smallskip}
N\,{\sc ii} ORLs$^{\mathrm{f}}$       & least-squares min.             & $      2100_{     -1800}^{+     78500}$ \\
\noalign{\smallskip}
\hline
\end{tabular}
\end{center}
\begin{list}{}{}
\item[$^{\mathrm{a}}$]Electron temperatures of low-excitation ions linked to $N_{\rm e}([$O~{\sc ii}]).
\item[$^{\mathrm{b}}$] $T_{\rm e}([$Ar~{\sc iii}]) linked with $N_{\rm e}([$Cl~{\sc iii}]), but $T_{\rm e}([$O~{\sc iii}]) with $N_{\rm e}([$Ar~{\sc iv}]) since $T_{\rm e}([$O~{\sc iii}$])> 25,000$\,K at $N_{\rm e} < 2 \times 10^5$\,cm$^{-3}$ (see Figure~\ref{ic4997:diagnostic:diagram}). 
\item[$^{\mathrm{c}}$]$\Delta T_{\rm +}$ the dichotomy between the \fnii\ and \hei\ temperatures, and $\Delta T_{\rm 2+}$ between the \fariii\ temperature and the mean temperature of the O~{\sc ii} and N~{\sc ii} ORLs. 
\item[$^{\mathrm{d}}$]$N_{\rm e}([$O~{\sc ii}]) coupled with $T_{\rm e}([$N~{\sc ii}]), $N_{\rm e}([$Cl~{\sc iii}]) with $T_{\rm e}([$Ar~{\sc iii}]), and $N_{\rm e}([$Ar~{\sc iv}]) with $T_{\rm e}([$O~{\sc iii}]).
\item[$^{\mathrm{e}}$] O\,{\sc ii} ORLs $\lambda$4119.2, 4153.3, 4319.6, 4349.4, 4416.9, 4609.4, 4610.2, 4641.8, 4649.1, 4650.8, 4661.6, and 4676.2\,{\AA}.
\item[$^{\mathrm{f}}$] N\,{\sc ii} ORLs $\lambda$4630.5, 5666.6, 5676.0, and 5679.6\,{\AA}.
\item[$^{\star}$]$T_{\rm e, rc}([$N~{\sc ii}]) obtained using the \fnii\ $\lambda$5755 line corrected for the recombination contamination. 
\end{list}
\end{table}

We derived the critical densities for temperature-diagnostic low-ionization lines at $T_{\rm e}=$\,18,000\,K as follows:
[N\,{\sc ii}] $\lambda\lambda$6584,\,5755, $N_{\rm cr}=9.97 \times 10^4$, $1.72\times 10^7$\,cm$^{-3}$; and 
[S\,{\sc ii}] $\lambda\lambda$6717,\,6731,\,4069,\,4076 $N_{\rm cr}=$\,2,290, 6,770, $3.24 \times 10^6$, $1.57 \times 10^6$ cm$^{-3}$; 
while for high-ionization $T_{\rm e}$-diagnostic lines:
[O\,{\sc iii}] $\lambda\lambda$4959,\,4363, $N_{\rm cr}=8.91 \times 10^5$, $3.61 \times 10^7$\,cm$^{-3}$ ($T_{\rm e}=$\,24,100\,K); and 
[Ar\,{\sc iii}] $\lambda\lambda$7751,\,5192, $N_{\rm cr}=5.75 \times 10^6$, $4.58 \times 10^7$\,cm$^{-3}$ ($T_{\rm e}=$\,14,300\,K), respectively. 
We see that the \fariii\ lines have the highest critical density among them, so they may be the most reliable lines in extremely dense ionized plasmas ($> 10^6$\, cm$^{-3}$). The \fnii\ and \foiii\ temperature-diagnostic lines may not be suitable for densities higher than $9 \times 10^4$ and $8 \times 10^5$\,cm$^{-3}$, respectively. Temperature-diagnostic lines emitted from a region having densities higher than the critical densities of those lines could yield a largely overestimated temperature, which can be seen in the value we obtained for the \foiii\ temperature. Assuming the \foiii\ lines are emitted from extremely dense regions ($> 5 \times 10^5$\,cm$^{-3}$), it could have a temperature below 20,000\,K according to our $N_{\rm e}$--$T_{\rm e}$  diagnostic diagram shown in Figure~\ref{ic4997:diagnostic:diagram}. The \fariii\ temperature, which has the highest critical density, is more suitable for this dense PN, and  therefore was used for higher ionized ions (X$^{i+}$, $i \geqslant 2$) in our first abundance analysis in Section~\ref{ic4997:abund}. In order to see the difference, we additionally assumed $T_{\rm e}$([O\,{\sc iii}]) for the higher ionized ions (X$^{i+}$, $i \geqslant 2$) in our second analysis. 

It can be seen that our physical conditions deduced from CELs are in agreement with the previous results. 
Our derived electron densities $\log N_{\rm e}($\fcliii$)\approx 4.5$ and $\log N_{\rm e}($\fariv$)\approx 5.4$ cm$^{-3}$ agree with  
$\log N_{\rm e}($\foii$) \sim 4.5$, $\log N_{\rm e}($Si\,{\sc iii}]$) \sim 5$, and $\log N_{\rm e}($C\,{\sc iii}]$) \sim 6.2$  \citep{Hyung1994}, and $\log N_{\rm e}\sim 5.6$ \citep{Aller1966,Ahern1978}. 
Furthermore, the electron temperatures $T_{\rm e,rc}($\fnii$)=$ 17,260\,K and $T_{\rm e}($\fariii$)=$ 14,180\,K that we calculated agree with $T_{\rm e} =$ 18,000 \,K \citep{Aller1966}, 15,700\,K \citep{Ahern1978}, 14,700\,K \citep{Flower1980}, and 15,000\,K \citep{Kiser1982}. 
Recently, \citet{Ruiz-Escobedo2022} also obtained $N_{\rm e} =31,700_{-11,000}^{+21,000}$ (low-excitation CELs) and $1.8_{-0.5}^{+0.9} \times 10^{6}$\,cm$^{-3}$ (high-excitation CELs), as well as $T_{\rm e} =16,200_{-2,800}^{+3,100}$\,K.

Using the effective recombination coefficients ($\alpha_{\rm eff}$) of ions, we implemented a least-squares minimization technique to determine the electron temperatures of heavy element ORLs according to \citet{McNabb2013} and \citet{Storey2013}. Additionally, we weighted our least-squares with flux uncertainties, so highly uncertain ORLs have lower weights in our least-squares minimization. We employed recombination coefficient to predict the theoretical ORL fluxes $I_{\rm predict}$ in the available temperature range and the ORL elemental abundances from \S\,\ref{ic4997:abund}. We used the dereddened fluxes $I_{\rm observe}$ and the predicted fluxes $I_{\rm predict}$  to calculate the weighted least-squares: $\chi^2 = \frac{1}{w_{\rm sum}} \sum_{j=1}^{N} w_{j}\big(I_{\rm observe}\left(\lambda_j\right) - I_{\rm predict}\left(\lambda_j\right)\big)^2$ in the full temperature range. The electron temperature $T_{\rm e}$ and density $N_{\rm e}$ are identified at the minimum value of the weighted least-squares (see Figure~\ref{fig:dd:orls}). The weights are calculated using the absolute errors of the dereddened fluxes, $w_{j}=1/\sigma^2_{I_{\rm observe}(\lambda_j)}$, with the total weight being ${w_{\rm sum}}=\sum_{j=1}^{N}w_{j}$. In the first iteration, we adopted the \fariii\ temperature, and we then iterated our calculations until there was no change in $T_{\rm e}$ and $N_{\rm e}$ identified at the minimum value of least-squares $\chi^2$. 

The logarithmic least-square distributions ($\log \chi^2$) calculated for \oii\ and \nii\ ORLs within the entire ranges of $T_{\rm e}$ and $N_{\rm e}$ are shown in Figure~\ref{fig:dd:orls}, with the minimum $\chi^2$ located at the crossing point of the two solid lines. Table~\ref{tab:result:cel:temp:dens} presents the electron temperatures and densities derived from the \oii\ and \nii\ ORLs using the least-squares minimization method. The confidence levels were obtained from the limits $\chi_{\rm min}^2 + 1$ 
of the least-squares normalized by $\chi_{\rm min}^2$ similar to \citet{Storey2013}. 
Although our electron temperature $T_{\rm e}($\oii$)= 7400_{-5300}^{+17400}$ determined from the \oii\ ORLs is highly uncertain, it may be more reliable than $T_{\rm e}($\nii$)= 6700_{-2900}^{+  8100}$ derived from the \nii\ ORLs 
that could be contaminated by fluorescence \citep{Escalante2005,Escalante2012}.
Our least-squares minimization \oii\ temperature, albeit with high uncertainties, agrees with $T_{\rm e}($\oii$)= 7100\pm 100$ found by \citet{Ruiz-Escobedo2022} using the empirical relationship of \citet{Peimbert2014}.
%We have more available \oii\ lines (than \nii) to constrain its electron temperature using the least-squares minimization. 
It can be seen that the ORL temperatures are much lower than those from the CELs.

Table~\ref{tab:result:cel:temp:dens} also lists the He\,{\sc i} temperature determined from the $\lambda5876/\lambda4472$ flux ratio using the analytic method of \citet{Benjamin1999} and the fitting parameters extended by \citet{Zhang2005a} to temperatures below 5000\,K. 
It can be seen that the helium temperature is below the \fnii\ temperature, but higher than the \nii\ and \oii\ temperatures. Our He\,{\sc i} $\lambda5876/\lambda4472$ temperature of $11070^{+1010}_{-880}$\,K agrees with the Balmer Jump temperature  $T_{\rm e}($BJ$)= 11300 \pm 700$\,K derived by \citet{Ruiz-Escobedo2022} using \citet{Liu2001}'s empirical method, as well as their He\,{\sc i} $\lambda$6678/$\lambda$5875 temperature of $10500 \pm 1100$\,K. However, \citet{Ruiz-Escobedo2022} also obtained $T_{\rm e}($He\,{\sc i}$)=8900^{+1100}_{-800}$\,K from the $\lambda7281/\lambda6678$ line ratio, which could be the most reliable He\,{\sc i} line ratio as argued by \citet{Zhang2005a} and \citet{Otsuka2010}. However, the NOT observations used in our study did not cover the He\,{\sc i} $\lambda$7281 line.

\section{Chemical Composition}
%\section{Abundance Analysis}
\label{ic4997:abund}

In the CEL analysis, we determined the ionic abundances of N, O, Ne, S, Cl, Ar and Fe from the dereddened fluxes of CELs and the physical conditions derived in \S\,\ref{ic4997:temp:dens} using the IDL program proEQUIB with
the Einstein spontaneous transition probabilities and the transition collision
strengths listed in Table~\ref{tab:atomicdata}. In our first approach, we assumed the temperature derived from the \foi\ flux ratio for O$^{0}$, the recombination-corrected \fnii\ temperature ($T_{\rm e,rc}$) for singly-ionized ions X$^+$, the \fariii\ temperature for X$^{i+}$ ions where $i \geqslant 2$, the density derived from the \foii\ flux ratio for neutral and singly-ionized ions, and the \fcliii\ density for X$^{i+}$ where $i \geqslant 2$. In our second approach, we adopted the same physical conditions used in the first one, apart from  the [O\,{\sc iii}] temperature and the [Ar\,{\sc iv}] density used for X$^{i+}$ where $i \geqslant 2$. The ionic abundances derived from CELs are presented in Table~\ref{tab:result:cel:abund}.
In our first analysis, our aim is to adopt the physical condition, $N_{\rm e}$([Cl\,{\sc iii}]$)=3.33 \times 10^{4}$\,cm$^{-3}$, for high-excitation lines that is less affected by the associated critical density limits. 
In our second analysis, $N_{\rm e}$([Ar\,{\sc iv}]$)=2.97 \times 10^{5}$\,cm$^{-3}$ can lead to the same ionic abundance O$^{2+}$ for the bright \foiii $\lambda$4959 line and faint \foiii $\lambda$4363 auroral line (see Table~\ref{tab:result:cel:abund}).
In our analyses, the IDL MCMC hammer was used for propagating flux measurement errors, excluding physical condition uncertainties, into our calculations to provide abundance uncertainties.

\begin{table*}
\caption{Ionic abundances derived from CELs. % for two different physical conditions. 
\label{tab:result:cel:abund}}
\footnotesize
\begin{center}
\begin{tabular}{llcccc}
\noalign{\smallskip}
\hline
\noalign{\smallskip}
 Ion  & Line & Weight\,$^{\mathrm{a}}$ & X${}^{i+}$/H${}^{+}$\,$^{\mathrm{a}}$ & Weight\,$^{\mathrm{b}}$ & X${}^{i+}$/H${}^{+}$\,$^{\mathrm{b}}$ \\
\noalign{\smallskip}
\hline
\noalign{\smallskip}
N$^{+}$    &[N~{\sc ii}]     $\lambda$5754.60  &     -- & [$ 1.073_{-0.017}^{+ 0.019} \times 10^{  -6}$] &    --  & [$ 1.073_{-0.017}^{+ 0.019} \times 10^{  -6}$] \\
\noalign{\smallskip}
N$^{+}$    &[N\,{\sc ii}]     $\lambda$6548.10  &    1 & $ 1.048_{-0.027}^{+ 0.028} \times 10^{  -6}$ &    1 & $ 1.048_{-0.027}^{+ 0.028} \times 10^{  -6}$ \\
\noalign{\smallskip}
N$^{+}$    &[N\,{\sc ii}]     $\lambda$6583.50  &    3 & $ 1.039_{-0.025}^{+ 0.026} \times 10^{  -6}$ &    3 & $ 1.039_{-0.025}^{+ 0.026} \times 10^{  -6}$ \\
\noalign{\smallskip}
O$^{0}$    &[O\,{\sc i}]      $\lambda$6300.34  &    3 & $ 3.136_{-0.060}^{+ 0.069} \times 10^{  -6}$ &    3 & $ 3.136_{-0.060}^{+ 0.069} \times 10^{  -6}$ \\
\noalign{\smallskip}
O$^{0}$    &[O\,{\sc i}]      $\lambda$6363.78  &    1 & $ 3.384_{-0.066}^{+ 0.073} \times 10^{  -6}$ &    1 & $ 3.384_{-0.066}^{+ 0.073} \times 10^{  -6}$ \\
\noalign{\smallskip}
O$^{+}$    &[O\,{\sc ii}]     $\lambda$3726.03  &    3 & $ 5.657_{-0.218}^{+ 0.207} \times 10^{  -6}$ &    3 & $ 5.657_{-0.218}^{+ 0.207} \times 10^{  -6}$ \\
\noalign{\smallskip}
O$^{+}$    &[O\,{\sc ii}]     $\lambda$3728.82  &    1 & $ 5.660_{-0.199}^{+ 0.194} \times 10^{  -6}$  &    1 & $ 5.660_{-0.199}^{+ 0.194} \times 10^{  -6}$ \\
\noalign{\smallskip}
O$^{2+}$   &[O~{\sc iii}]    $\lambda$4363.21  &    --  & [$ 3.501_{-0.070}^{+ 0.058} \times 10^{  -4}$] &     -- & [$ 3.507_{-0.067}^{+ 0.063} \times 10^{  -5}$] \\
\noalign{\smallskip}
O$^{2+}$   &[O\,{\sc iii}]    $\lambda$4958.91  &    1 & $ 8.140_{-0.093}^{+ 0.093} \times 10^{  -5}$ &    1 & $ 3.508_{-0.040}^{+ 0.041} \times 10^{  -5}$ \\
\noalign{\smallskip}
Ne$^{2+}$  &[Ne\,{\sc iii}]   $\lambda$3868.75  &    3 & $ 4.008_{-0.090}^{+ 0.075} \times 10^{  -5}$ &   3  & $ 1.131_{-0.025}^{+ 0.024} \times 10^{  -5}$ \\
\noalign{\smallskip}
Ne$^{2+}$  &[Ne\,{\sc iii}]   $\lambda$3967.46  &    1 & $ 3.822_{-0.103}^{+ 0.085} \times 10^{  -5}$ &    1 & $ 1.079_{-0.028}^{+ 0.023} \times 10^{  -5}$ \\
\noalign{\smallskip}
S$^{+}$    &[S\,{\sc ii}]     $\lambda$4068.60  &    3 & $ 5.514_{-0.170}^{+ 0.153} \times 10^{  -8}$ &   4  & $ 5.514_{-0.170}^{+ 0.153} \times 10^{  -8}$ \\
\noalign{\smallskip}
S$^{+}$    &[S\,{\sc ii}]     $\lambda$4076.35  &    1 & $ 5.391_{-0.226}^{+ 0.200} \times 10^{  -8}$ &    1 & $ 5.391_{-0.226}^{+ 0.200} \times 10^{  -8}$ \\
\noalign{\smallskip}
S$^{+}$    &[S\,{\sc ii}]     $\lambda$6716.44  &    1 & $ 7.985_{-0.297}^{+ 0.313} \times 10^{  -8}$ &    1 & $ 7.985_{-0.297}^{+ 0.313} \times 10^{  -8}$ \\
\noalign{\smallskip}
S$^{+}$    &[S\,{\sc ii}]     $\lambda$6730.82  &    2 & $ 4.201_{-0.120}^{+ 0.133} \times 10^{  -8}$ &    2 & $ 4.201_{-0.120}^{+ 0.133} \times 10^{  -8}$ \\
\noalign{\smallskip}
S$^{2+}$   &[S\,{\sc iii}]    $\lambda$6312.10  &    1 & $ 1.212_{-0.025}^{+ 0.025} \times 10^{  -6}$ &    1 & $ 2.926_{-0.058}^{+ 0.070} \times 10^{  -7}$ \\
\noalign{\smallskip}
Cl$^{2+}$  &[Cl\,{\sc iii}]   $\lambda$5517.66  &    1 & $ 1.043_{-0.023}^{+ 0.025} \times 10^{  -8}$ &    1 & $ 2.243_{-0.050}^{+ 0.051} \times 10^{  -8}$ \\
\noalign{\smallskip}
Cl$^{2+}$  &[Cl\,{\sc iii}]   $\lambda$5537.60  &    3 & $ 1.044_{-0.019}^{+ 0.021} \times 10^{  -8}$ &    5 & $ 1.487_{-0.028}^{+ 0.030} \times 10^{  -8}$ \\
\noalign{\smallskip}
Ar$^{2+}$  &[Ar\,{\sc iii}]   $\lambda$5191.82  &    1 & $ 3.300_{-0.073}^{+ 0.073} \times 10^{  -7}$ &    1 & $ 6.758_{-0.148}^{+ 0.158} \times 10^{  -8}$ \\
\noalign{\smallskip}
Ar$^{2+}$  &[Ar\,{\sc iii}]   $\lambda$7135.80  &   58 & $ 3.301_{-0.091}^{+ 0.099} \times 10^{  -7}$ &   24 & $ 1.576_{-0.043}^{+ 0.046} \times 10^{  -7}$ \\
\noalign{\smallskip}
Ar$^{3+}$  &[Ar\,{\sc iv}]    $\lambda$4711.37  &   13 & $ 2.329_{-0.082}^{+ 0.071} \times 10^{  -8}$ &    2 & $ 2.569_{-0.088}^{+ 0.079} \times 10^{  -8}$ \\
\noalign{\smallskip}
Ar$^{3+}$  &[Ar\,{\sc iv}]    $\lambda$4740.17  &   36 & $ 5.437_{-0.060}^{+ 0.051} \times 10^{  -8}$ &   10 & $ 2.569_{-0.027}^{+ 0.026} \times 10^{  -8}$ \\
\noalign{\smallskip}
Ar$^{3+}$  &[Ar\,{\sc iv}]    $\lambda$7237.26  &    1 & $ 2.973_{-0.210}^{+ 0.211} \times 10^{  -7}$ &    1 & $ 3.665_{-0.252}^{+ 0.258} \times 10^{  -8}$ \\
\noalign{\smallskip}
Ar$^{3+}$  &[Ar\,{\sc iv}]    $\lambda$7262.76  &    1 & $ 3.534_{-0.141}^{+ 0.143} \times 10^{  -7}$ &    1 & $ 4.359_{-0.176}^{+ 0.178} \times 10^{  -8}$ \\
\noalign{\smallskip}
Fe$^{2+}$  &[Fe\,{\sc iii}]   $\lambda$4701.62  &    2 & $ 6.848_{-0.157}^{+ 0.141} \times 10^{  -7}$ &    1 & $ 2.223_{-0.050}^{+ 0.047} \times 10^{  -7}$ \\
\noalign{\smallskip}
Fe$^{2+}$  &[Fe\,{\sc iii}]   $\lambda$4733.91  &    1 & $ 4.773_{-0.245}^{+ 0.234} \times 10^{  -7}$ &    1 & $ 1.467_{-0.077}^{+ 0.072} \times 10^{  -7}$ \\
\noalign{\smallskip}
Fe$^{2+}$  &[Fe\,{\sc iii}]   $\lambda$4769.40  &    5 & $ 9.479_{-0.334}^{+ 0.294} \times 10^{  -8}$ &    5 & $ 3.077_{-0.108}^{+ 0.101} \times 10^{  -8}$ \\
\noalign{\smallskip}
Fe$^{2+}$  &[Fe\,{\sc iii}]   $\lambda$5270.40  &    6 & $ 1.502_{-0.064}^{+ 0.063} \times 10^{  -7}$ &    6 & $ 5.303_{-0.229}^{+ 0.232} \times 10^{  -8}$ \\
\noalign{\smallskip}
\hline
\end{tabular}
\end{center}
\begin{list}{}{}
\item[$^{\mathrm{a}}$]$T_{\rm e}$([O\,{\sc i}]), $T_{\rm e,rc}$([N\,{\sc ii}]), and $T_{\rm e}$([Ar\,{\sc iii}]) adopted for X$^0$, X$^+$ and X$^{i+}$ ($i \geqslant 2$), respectively; and $N_{\rm e}$([O\,{\sc ii}]) and $N_{\rm e}$([Cl\,{\sc iii}]) assumed for X$^{i+}$ ($i=0,1$) and X$^{i+}$ ($i \geqslant 2$), respectively.
\item[$^{\mathrm{b}}$]The same as the physical conditions used in the first approach ($\mathrm{a}$), except for $T_{\rm e}$([O\,{\sc iii}]) and $N_{\rm e}$([Ar\,{\sc iv}]) used for X$^{i+}$ ($i \geqslant 2$).
\end{list}
\begin{tablenotes}
\item[1]\textbf{Note.} The weights, which are calculated based on the predicted intrinsic fluxes, are used to calculate the average value of each ionic abundance presented in Table~\ref{tab:result:total:abund}. 
The average ionic abundances and their uncertainties are calculated according to the weights of the ionic abundances, and the values in square brackets $[~]$ are not used for the average ionic abundances listed in Table~\ref{tab:result:total:abund}.
\end{tablenotes}
\end{table*}

\begin{table}
\caption{Ionic abundances derived from ORLs. \label{tab:result:orl:abund}}
\begin{center}
\footnotesize
\begin{tabular}{llcc}
\hline
\noalign{\smallskip}
 Ion  & Line & Weight & X${}^{i+}$/H${}^{+}$\,$^{\mathrm{a}}$  \\
\noalign{\smallskip}
\hline
\noalign{\smallskip}
He$^{+}$   &He~{\sc i}       $\lambda$4120.84  &    1 & $ 1.318_{-0.041}^{+ 0.040} \times 10^{  -2}$ \\
\noalign{\smallskip}
He$^{+}$   &He~{\sc i}       $\lambda$4387.93  &    3 & $ 1.063_{-0.013}^{+ 0.010} \times 10^{  -1}$ \\
\noalign{\smallskip}
He$^{+}$   &He~{\sc i}       $\lambda$4471.50  &   21 & $ 1.155_{-0.007}^{+ 0.006} \times 10^{  -1}$ \\
\noalign{\smallskip}
He$^{+}$   &He~{\sc i}       $\lambda$5047.74  &    1 & $ 1.174_{-0.054}^{+ 0.047} \times 10^{  -1}$ \\
\noalign{\smallskip}
He$^{+}$   &He~{\sc i}       $\lambda$5875.66  &   65 & $ 9.763_{-0.085}^{+ 0.107} \times 10^{  -2}$ \\
\noalign{\smallskip}
He$^{+}$   &He~{\sc i}       $\lambda$6678.16  &   18 & $ 8.965_{-0.137}^{+ 0.119} \times 10^{  -2}$ \\
\noalign{\smallskip}
He$^{+}$   &He~{\sc i}       $\lambda$7065.25  &   22 & $ 1.427_{-0.027}^{+ 0.025} \times 10^{  -1}$ \\
\noalign{\smallskip}
C$^{2+}$   &C~{\sc ii}       $\lambda$6151.43  &    1 & $ 6.767_{-0.361}^{+ 0.318} \times 10^{  -5}$ \\
\noalign{\smallskip}
C$^{3+}$   &C~{\sc iii}      $\lambda$4647.42  &    2 & $ 1.601_{-0.027}^{+ 0.029} \times 10^{  -4}$ \\
\noalign{\smallskip}
C$^{3+}$   &C~{\sc iii}      $\lambda$4650.25  &    1 & $ 1.533_{-0.091}^{+ 0.079} \times 10^{  -4}$ \\
\noalign{\smallskip}
N$^{2+}$   &N~{\sc ii}       $\lambda$4630.54  &    3 & $ 1.716_{-0.022}^{+ 0.020} \times 10^{  -4}$ \\
\noalign{\smallskip}
N$^{2+}$   &N~{\sc ii}       $\lambda$5666.63  &    2 & $ 7.717_{-0.156}^{+ 0.126} \times 10^{  -5}$ \\
\noalign{\smallskip}
N$^{2+}$   &N~{\sc ii}       $\lambda$5676.02  &    1 & $ 5.659_{-0.214}^{+ 0.251} \times 10^{  -5}$ \\
\noalign{\smallskip}
N$^{2+}$   &N~{\sc ii}       $\lambda$5679.56  &    5 & $ 7.349_{-0.083}^{+ 0.079} \times 10^{  -5}$ \\
\noalign{\smallskip}
N$^{3+}$   &N~{\sc iii}      $\lambda$4640.64  &    1 & $ 3.975_{-0.150}^{+ 0.149} \times 10^{  -5}$ \\
\noalign{\smallskip}
O$^{2+}$   &O~{\sc ii}       $\lambda$4119.22  &    7 & $ 6.962_{-0.167}^{+ 0.137} \times 10^{  -4}$ \\
\noalign{\smallskip}
O$^{2+}$   &O~{\sc ii}       $\lambda$4153.30  &    6 & $ 7.777_{-0.302}^{+ 0.281} \times 10^{  -4}$ \\
\noalign{\smallskip}
O$^{2+}$   &O~{\sc ii}       $\lambda$4319.63  &    3 & $ 1.040_{-0.057}^{+ 0.042} \times 10^{  -3}$ \\
\noalign{\smallskip}
O$^{2+}$   &O~{\sc ii}       $\lambda$4349.43  &    9 & $ 7.608_{-0.171}^{+ 0.178} \times 10^{  -4}$ \\
\noalign{\smallskip}
O$^{2+}$   &O~{\sc ii}       $\lambda$4416.97  &    4 & $ 7.179_{-0.498}^{+ 0.415} \times 10^{  -4}$ \\
\noalign{\smallskip}
O$^{2+}$   &O~{\sc ii}       $\lambda$4609.44  &    3 & $ 6.261_{-0.178}^{+ 0.145} \times 10^{  -4}$ \\
\noalign{\smallskip}
O$^{2+}$   &O~{\sc ii}       $\lambda$4610.20  &    1 & $ 1.013_{-0.052}^{+ 0.043} \times 10^{  -3}$ \\
\noalign{\smallskip}
O$^{2+}$   &O~{\sc ii}       $\lambda$4641.81  &   20 & $ 7.719_{-0.166}^{+ 0.167} \times 10^{  -4}$ \\
\noalign{\smallskip}
O$^{2+}$   &O~{\sc ii}       $\lambda$4649.13  &   36 & $ 4.682_{-0.084}^{+ 0.072} \times 10^{  -4}$ \\
\noalign{\smallskip}
O$^{2+}$   &O~{\sc ii}       $\lambda$4650.84  &    9 & $ 5.114_{-0.235}^{+ 0.197} \times 10^{  -4}$ \\
\noalign{\smallskip}
O$^{2+}$   &O~{\sc ii}       $\lambda$4661.63  &   10 & $ 9.617_{-0.115}^{+ 0.080} \times 10^{  -4}$ \\
\noalign{\smallskip}
O$^{2+}$   &O~{\sc ii}       $\lambda$4676.23  &    7 & $ 4.066_{-0.311}^{+ 0.276} \times 10^{  -4}$ \\
\noalign{\smallskip}
\hline
\end{tabular}
\begin{list}{}{}
\item[$^{\mathrm{a}}$]$T_{\rm e}$(He\,{\sc i}) and $N_{\rm e}$([O\,{\sc ii}]) adopted for He$^{+}$; and $T_{\rm e}$([Ar\,{\sc iii}]) and $N_{\rm e}$([Cl\,{\sc iii}]) assumed for X$^{i+}$ ($i \geqslant 2$).
\end{list}
\end{center}
\small
\begin{tablenotes}
\item[1]\textbf{Note.} The weights calculated according to the theoretical fluxes are used to obtain the average ionic abundances listed in Table~\ref{tab:result:total:abund}. The average ionic abundances and their uncertainties presented in Table~\ref{tab:result:total:abund} largely depend on the ORLs with higher weights. 
\end{tablenotes}
\end{table}

\begin{table*}
\footnotesize
\caption{Averaged ionic abundances and elemental abundances. \label{tab:result:total:abund}}
\begin{center}
\begin{tabular}{lclcc}
\noalign{\smallskip}
\hline
\noalign{\smallskip}
 Ion  & Typ. & {\it icf} ref. & X/H\,$^{\mathrm{a}}$ & X/H\,$^{\mathrm{b}}$ \\
\noalign{\smallskip}
\hline
\noalign{\smallskip}
He$^{+}$/H       & ORL     &          & $       1.067_{      -0.008}^{+       0.008} \times 10^{  -1}$ & $       1.067_{      -0.008}^{+       0.008} \times 10^{  -1}$\\
\noalign{\smallskip}
He/H             & ORL     &          & $       1.067_{      -0.008}^{+       0.008} \times 10^{  -1}$ & $       1.067_{      -0.008}^{+       0.008} \times 10^{  -1}$\\
\noalign{\smallskip}
\noalign{\medskip}
C$^{2+}$/H       & ORL     &          & $       6.767_{      -0.338}^{+       0.348} \times 10^{  -5}$ & $       6.767_{      -0.338}^{+       0.348} \times 10^{  -5}$\\
\noalign{\smallskip}
C$^{3+}$/H       & ORL     &          & $       1.578_{      -0.040}^{+       0.039} \times 10^{  -4}$ & $       1.578_{      -0.040}^{+       0.039} \times 10^{  -4}$\\
\noalign{\smallskip}
{\it icf}(C)     & ORL     & WL07     & $       1.070_{      -0.003}^{+       0.002}$ & $       1.161_{      -0.007}^{+       0.008}$ \\
\noalign{\smallskip}
C/H              & ORL     & WL07     & $       2.412_{      -0.080}^{+       0.063} \times 10^{  -4}$ & $       2.619_{      -0.073}^{+       0.104} \times 10^{  -4}$ \\
\noalign{\smallskip}
\noalign{\medskip}
{\it icf}(C)     & ORL     & DMS14    & $       1.662_{      -0.388}^{+       0.420}$ & $       1.604_{      -0.381}^{+       0.305}$ \\
\noalign{\smallskip}
C/H              & ORL     & DMS14    & $       1.125_{      -0.321}^{+       0.343} \times 10^{  -4}$ & $       1.085_{      -0.262}^{+       0.276} \times 10^{  -4}$ \\
\noalign{\smallskip}
\noalign{\medskip}
N$^{2+}$/H       & ORL     &          & $       9.938_{      -0.109}^{+       0.093} \times 10^{  -5}$ & $       9.938_{      -0.109}^{+       0.093} \times 10^{  -5}$\\
\noalign{\smallskip}
N$^{3+}$/H       & ORL     &          & $       3.975_{      -0.150}^{+       0.149} \times 10^{  -5}$ & $       3.975_{      -0.150}^{+       0.149} \times 10^{  -5}$\\
\noalign{\smallskip}
{\it icf}(N)     & ORL     & WL07     & $       1.070_{      -0.003}^{+       0.002}$ & $       1.161_{      -0.008}^{+       0.011}$ \\
\noalign{\smallskip}
N/H              & ORL     & WL07     & $       1.488_{      -0.029}^{+       0.025} \times 10^{  -4}$ & $       1.616_{      -0.031}^{+       0.037} \times 10^{  -4}$ \\
\noalign{\smallskip}
\noalign{\medskip}
O$^{2+}$/H       & ORL     &          & $       6.489_{      -0.082}^{+       0.055} \times 10^{  -4}$ & $       6.489_{      -0.074}^{+       0.068} \times 10^{  -4}$ \\
\noalign{\smallskip}
{\it icf}(O)     & ORL     & WL07     & $       1.070_{      -0.010}^{+       0.010}$ & $       1.161_{      -0.012}^{+       0.012}$ \\
\noalign{\smallskip}
O/H              & ORL     & WL07     & $       6.940_{      -0.123}^{+       0.092} \times 10^{  -4}$ & $       7.536_{      -0.141}^{+       0.125} \times 10^{  -4}$ \\
\noalign{\smallskip}
\noalign{\medskip}
N$^{+}$/H        & CEL     &          & $       1.041_{      -0.024}^{+       0.026} \times 10^{  -6}$ & $       1.041_{      -0.024}^{+       0.026} \times 10^{  -6}$ \\
\noalign{\smallskip}
{\it icf}(N)     & CEL     & KB94     & $      15.388_{      -0.424}^{+       0.823}$ & $       7.201_{      -0.346}^{+       0.328}$ \\
\noalign{\smallskip}
N/H              & CEL     & KB94     & $       1.602_{      -0.069}^{+       0.097} \times 10^{  -5}$ & $       7.496_{      -0.411}^{+       0.501} \times 10^{  -6}$ \\
\noalign{\smallskip}
\noalign{\medskip}
{\it icf}(N)     & CEL     & DMS14    & $      61.037_{     -21.027}^{+      32.963}$ & $      25.616_{      -8.494}^{+      12.624}$ \\
\noalign{\smallskip}
N/H              & CEL     & DMS14    & $       6.354_{      -2.686}^{+       3.780} \times 10^{  -5}$ & $       2.667_{      -1.134}^{+       1.430} \times 10^{  -5}$ \\
\noalign{\smallskip}
\noalign{\medskip}
O$^{0}$/H        & CEL     &          & $       3.198_{      -0.054}^{+       0.059} \times 10^{  -6}$ & $       3.198_{      -0.054}^{+       0.059} \times 10^{  -6}$ \\
\noalign{\smallskip}
O$^{+}$/H        & CEL     &          & $       5.657_{      -0.189}^{+       0.134} \times 10^{  -6}$ & $       5.657_{      -0.189}^{+       0.134} \times 10^{  -6}$ \\
\noalign{\smallskip}
O$^{2+}$/H       & CEL     &          & $       8.140_{      -0.093}^{+       0.093} \times 10^{  -5}$ & $       3.508_{      -0.040}^{+       0.041} \times 10^{  -5}$ \\
\noalign{\smallskip}
{\it icf}(O)     & CEL     & KB94     & $       1.000_{      -0.008}^{+       0.010}$ & $       1.000_{      -0.011}^{+       0.009}$ \\
\noalign{\smallskip}
O/H              & CEL     & KB94     & $       8.706_{      -0.131}^{+       0.135} \times 10^{  -5}$  & $       4.074_{      -0.079}^{+       0.072} \times 10^{  -5}$ \\
\noalign{\smallskip}
\noalign{\medskip}
{\it icf}(O)     & CEL     & DMS14    & $       1.000_{      -0.030}^{+       0.030}$ & $       1.000_{      -0.030}^{+       0.030}$ \\
\noalign{\smallskip}
O/H              & CEL     & DMS14    & $       8.706_{      -0.305}^{+       0.342} \times 10^{  -5}$ & $       4.074_{      -0.148}^{+       0.165} \times 10^{  -5}$ \\
\noalign{\smallskip}
\noalign{\medskip}
\hline
\end{tabular}
\begin{list}{}{}
\item[$^{\mathrm{a}}$]$T_{\rm e}$([O\,{\sc i}]), $T_{\rm e,rc}$([N\,{\sc ii}]), and $T_{\rm e}$([Ar\,{\sc iii}]) adopted for CEL X$^0$, X$^+$ and X$^{i+}$ ($i \geqslant 2$), respectively; $N_{\rm e}$([O\,{\sc ii}]) and $N_{\rm e}$([Cl\,{\sc iii}]) assumed for CEL X$^{i+}$ ($i=0,1$) and X$^{i+}$ ($i \geqslant 2$), respectively; $T_{\rm e}$(He\,{\sc i}]) and $N_{\rm e}$([O\,{\sc ii}]) adopted for He$^{+}$; and $T_{\rm e}$([Ar\,{\sc iii}]) and $N_{\rm e}$([Cl\,{\sc iii}]) assumed for ORL X$^{i+}$ ($i \geqslant 2$).
\item[$^{\mathrm{b}}$]The same as the physical conditions used in the first approach ($\mathrm{a}$), except for $T_{\rm e}$([O\,{\sc iii}]) and $N_{\rm e}$([Ar\,{\sc iv}]) used for CEL X$^{i+}$ ($i \geqslant 2$).
\end{list}
\end{center}
\begin{tablenotes}
\item[1]\textbf{Note.} The second column indicates that the ionic abundances were obtained from CELs or ORLs. 
The third column gives the reference of the \textit{icf} methods used for calculating elemental abundances. References are as follows: 
DMS14 -- \citet{Delgado-Inglada2014}; 
ITL94 -- \citet{Izotov1994}; 
KB94 -- \citet{Kingsburgh1994}; 
L00 -- \citet{Liu2000}; 
WL07 -- \citet{Wang2007}. 
\end{tablenotes}
\end{table*}

\setcounter{table}{8}
\begin{table*}
%\caption{-- \textit{continued}}
\contcaption{}
\footnotesize
%\contcaption{}
%\centering
\begin{center}
\begin{tabular}{lclcc}
\noalign{\smallskip}
\hline
\noalign{\smallskip}
 Ion  & Typ. & {\it icf} ref. & X/H\,$^{\mathrm{a}}$ & X/H\,$^{\mathrm{b}}$ \\
\noalign{\smallskip}
\hline
\noalign{\smallskip}
Ne$^{2+}$/H      & CEL     &          & $       3.961_{      -0.081}^{+       0.051} \times 10^{  -5}$  & $       1.118_{      -0.025}^{+       0.021} \times 10^{  -5}$ \\
\noalign{\smallskip}
{\it icf}(Ne)    & CEL     & KB94     & $       1.070_{      -0.023}^{+       0.029}$ & $       1.161_{      -0.037}^{+       0.027}$ \\
\noalign{\smallskip}
Ne/H             & CEL     & KB94     & $       4.237_{      -0.133}^{+       0.139} \times 10^{  -5}$ & $       1.298_{      -0.052}^{+       0.047} \times 10^{  -5}$ \\
\noalign{\smallskip}
\noalign{\medskip}
{\it icf}(Ne)    & CEL     & DMS14    & $       1.551_{      -0.331}^{+       0.271}$ & $       1.889_{      -0.411}^{+       0.319}$ \\
\noalign{\smallskip}
Ne/H             & CEL     & DMS14    & $       6.142_{      -1.526}^{+       1.376} \times 10^{  -5}$ & $       2.111_{      -0.509}^{+       0.405} \times 10^{  -5}$ \\
\noalign{\smallskip}
\noalign{\medskip}
S$^{+}$/H        & CEL     &          & $       5.474_{      -0.124}^{+       0.114} \times 10^{  -8}$ & $       5.474_{      -0.124}^{+       0.114} \times 10^{  -8}$ \\
\noalign{\smallskip}
S$^{2+}$/H       & CEL     &          & $       1.212_{      -0.025}^{+       0.025} \times 10^{  -6}$ & $       2.926_{      -0.058}^{+       0.070} \times 10^{  -7}$ \\
\noalign{\smallskip}
{\it icf}(S)     & CEL     & KB94     & $       1.763_{      -0.014}^{+       0.028}$ & $       1.404_{      -0.021}^{+       0.017}$ \\
\noalign{\smallskip}
S/H              & CEL     & KB94     & $       2.234_{      -0.055}^{+       0.073} \times 10^{  -6}$ & $       4.877_{      -0.117}^{+       0.153} \times 10^{  -7}$ \\
\noalign{\smallskip}
\noalign{\medskip}
{\it icf}(S)     & CEL     & DMS14    & $       1.733_{      -0.425}^{+       0.334}$ & $       1.347_{      -0.345}^{+       0.246}$ \\
\noalign{\smallskip}
S/H              & CEL     & DMS14    & $       2.195_{      -0.616}^{+       0.544} \times 10^{  -6}$ & $       4.680_{      -1.407}^{+       1.063} \times 10^{  -7}$ \\
\noalign{\smallskip}
\noalign{\medskip}
Cl$^{2+}$/H      & CEL     &          & $       1.043_{      -0.017}^{+       0.020} \times 10^{  -8}$ & $       1.613_{      -0.029}^{+       0.030} \times 10^{  -8}$ \\
\noalign{\smallskip}
{\it icf}(Cl)    & CEL     & L00      & $       1.842_{      -0.070}^{+       0.103}$ & $       1.666_{      -0.080}^{+       0.069}$ \\
\noalign{\smallskip}
Cl/H             & CEL     & L00      & $       1.922_{      -0.085}^{+       0.137} \times 10^{  -8}$ & $       2.688_{      -0.152}^{+       0.155} \times 10^{  -8}$ \\
\noalign{\smallskip}
\noalign{\medskip}
{\it icf}(Cl)    & CEL     & DMS14    & $      15.388_{      -2.697}^{+       1.116}$ & $       7.201_{      -1.340}^{+       0.448}$ \\
\noalign{\smallskip}
Cl/H             & CEL     & DMS14    & $       1.606_{      -0.303}^{+       0.186} \times 10^{  -7}$  & $       1.162_{      -0.228}^{+       0.109} \times 10^{  -7}$ \\
\noalign{\smallskip}
\noalign{\medskip}
Ar$^{2+}$/H      & CEL     &          & $       3.301_{      -0.111}^{+       0.122} \times 10^{  -7}$ & $       1.542_{      -0.050}^{+       0.051} \times 10^{  -7}$ \\
\noalign{\smallskip}
Ar$^{3+}$/H      & CEL     &          & $       5.713_{      -0.098}^{+       0.075} \times 10^{  -8}$ & $       2.810_{      -0.042}^{+       0.038} \times 10^{  -8}$ \\
\noalign{\smallskip}
{\it icf}(Ar)    & CEL     & KB94     & $       1.070_{      -0.006}^{+       0.005}$ & $       1.161_{      -0.014}^{+       0.019}$ \\
\noalign{\smallskip}
Ar/H             & CEL     & KB94     & $       4.141_{      -0.155}^{+       0.125} \times 10^{  -7}$ & $       2.116_{      -0.090}^{+       0.098} \times 10^{  -7}$ \\
\noalign{\smallskip}
\noalign{\medskip}
{\it icf}(Ar)    & CEL     & DMS14    & $       1.530_{      -1.139}^{+       0.992}$ & $       1.356_{      -1.014}^{+       0.884}$ \\
\noalign{\smallskip}
Ar/H             & CEL     & DMS14    & $       5.051_{      -4.692}^{+       4.192} \times 10^{  -7}$ & $       2.091_{      -2.010}^{+       1.525} \times 10^{  -7}$ \\
\noalign{\smallskip}
\noalign{\medskip}
Fe$^{2+}$/H      & CEL     &          & $       2.301_{      -0.058}^{+       0.050} \times 10^{  -7}$ & $       6.469_{      -0.169}^{+       0.144} \times 10^{  -8}$ \\
\noalign{\smallskip}
{\it icf}(Fe)    & CEL     & ITL94    & $      19.235_{      -0.525}^{+       0.883}$ & $       9.001_{      -0.489}^{+       0.415}$ \\
\noalign{\smallskip}
Fe/H             & CEL     & ITL94    & $       4.427_{      -0.183}^{+       0.280} \times 10^{  -6}$ & $       5.823_{      -0.384}^{+       0.421} \times 10^{  -7}$ \\
\noalign{\smallskip}
\noalign{\medskip}
{\it adf}(N$^{2+}$)       & ORL/CEL &          & $       6.635_{      -0.447}^{+       0.415}$ & $      15.396_{      -1.109}^{+       0.939}$ \\
\noalign{\smallskip}
{\it adf}(O$^{2+}$)       & ORL/CEL &          & $       7.972_{      -0.171}^{+       0.148}$ & $      18.499_{      -0.352}^{+       0.438}$ \\
\noalign{\smallskip}
\hline
\end{tabular}
\end{center}
\end{table*}

In the ORL analysis, ionic species of He, C, N, and O were obtained from the dereddened fluxes of ORLs. The recombination coefficient of each line was calculated by interpolating the given physical conditions 
on the two-dimensional $T_{\rm e}$--$N_{\rm e}$ grids of recombination atomic data listed in Table~\ref{tab:atomicdata}. 
We assumed the He\,{\sc i} temperature and the [O\,{\sc ii}] density to calculate the ionic abundance He$^{+}$. The temperature derived from the \fariii\ lines and the density from the \fcliii\ lines were adopted with highly-ionized ions X$^{i+}$ ($i \geqslant 2$) in our ORL calculations. 
We note that the recombination atomic data of most ORLs have not yet been computed at densities higher than $10^5$\,cm$^{-3}$. The uncertainties of our derived ionic abundances were estimated by propagating the flux errors in the calculations with the MCMC hammer. The ionic abundances derived from ORLs are presented in Table~\ref{tab:result:orl:abund}. 
It can be seen that the average ionic abundances from the ORLs in Table~\ref{tab:result:total:abund} are in reasonable agreement with the previous results.
We derived 
$N($O$^{2+}$)/$N($H$^{+})= 6.5 \times 10^{-4}$ that roughly agrees with 
$N($O$^{2+}$)/$N($H$^{+})= 10.0 \times 10^{-4}$ (case A) and $N($O$^{2+}$)/$N($H$^{+})= 7.1 \times 10^{-4}$ (case B)
by \citet[][]{Hyung1994}. 
However, we obtained $N($N$^{2+}$)/$N($H$^{+})= 1.0 \times 10^{-4}$ and $N($C$^{2+}$)/$N($H$^{+})= 6.8 \times 10^{-5}$, 
which are around two and four times lower than $N($N$^{2+}$)/$N($H$^{+})= 2.3 \times 10^{-4}$ and
$N($C$^{2+}$)/$N($H$^{+})= 2.4 \times 10^{-4}$ found by \citet[][]{Hyung1994}. 

We determined elemental abundances from ionic abundances using two different sets of the ionization correction factors (\textit{icf}). In our first (standard) \textit{icf} approach, we employed the \textit{icf} formulas of \citet[][KB94]{Kingsburgh1994}, the chlorine \textit{icf} method of \citet[][L00]{Liu2000}, and the iron \textit{icf} formula of \citet[][ITL94]{Izotov1994} for the CEL abundance analysis, and the \textit{icf} formulas from \citet[][WL07]{Wang2007} for the ORL abundance analysis. In our second \textit{icf} approach, we obtained elemental abundances using the schemes and uncertainty methods of \citet[][DMS14]{Delgado-Inglada2014}. Table~\ref{tab:result:total:abund} presents the elemental abundances based on the different \textit{icf} formulas and the two different assumptions of the physical conditions.

For the PN IC\,4997, we obtained the ORL/CEL abundance discrepancy factors (ADFs\,$\equiv$\,ORLs/CELs) of the weighted-average ionic abundances O$^{2+}$ and N$^{2+}$ from the \oii\ and \nii\ ORLs and the \foiii\ and \fnii\ CELs. The CEL N$^{2+}$ ionic abundance is obtained using the general assumption of N$^{2+}$\,=\,N$^+$\,$\times$\,O$^{2+}$/O$^+$ \citep{Kingsburgh1994}. From Table~\ref{tab:result:total:abund}, it can be seen that ADF(O$^{2+}) \gtrsim 8 $ and ADF(N$^{2+}) \gtrsim 7 $.
We determined the CEL--ORL temperature dichotomy $\Delta T_{\rm +}$ between the \fnii\ and \hei\ temperatures associated with the singly-ionized ions N$^{+}$ and He$^{+}$, and $\Delta T_{\rm 2+}$ between the \fariii\ temperature and the mean temperature of \oii\ and \nii\ ORLs corresponding  to the doubly-ionized ions Ar$^{2+}$, O$^{2+}$, and N$^{2+}$. 
Accordingly, the temperature dichotomies are $\Delta T_{\rm +} \equiv
T_{\rm e}(\textrm{[N\,{\sc ii}]}) - T_{\rm e}(\textrm{He\,{\sc i}}) = 7410_{-1380}^{+1470}$\,K, and $\Delta T_{\rm 2+} \equiv
T_{\rm e}(\textrm{[Ar\,{\sc iii}]}) - T_{\rm e}\langle\textrm{O\,{\sc ii}},\textrm{N\,{\sc ii}}\rangle = 7210_{-7210}^{+4500}$\,K.
Furthermore, \citet{Ruiz-Escobedo2022} derived ADF(O$^{2+}) \approx 5 $, while their plasma diagnostics also imply
the dichotomies $T_{\rm e}(\textrm{CELs}) - T_{\rm e}(\textrm{BJ}) = 4900$\,K and
$T_{\rm e}(\textrm{CELs}) - T_{\rm e}(\textrm{He\,{\sc i} 7281/6678}) = 7300$\,K  where $T_{\rm e}(\textrm{CELs})=16200$\,K.

Taking the oxygen abundance as a metallicity indicator, we see that the PN IC\,4997 is metal-poor with [O/H]~=~$-0.75$ (or $-1.08$) from the CEL abundance analysis using the standard \textit{icf} formulas with the first (or second) assumptions of physical conditions. We assumed the solar composition of \citet{Asplund2009}. As sulfur, argon, and chlorine also remain untouched in low- to intermediate-mass stars, they may be reliable indicators of metallicity. From our CEL abundance analyses, we found [S/H]~=~$-0.77$ in our first approach (or $-1.43$ in our second one), [Ar/H]~=~$-0.70$ (or $-1.08$), [Cl/H]~=~$-1.22$ (or $-1.07$), and [Fe/H]~=~$-0.85$ (or $-1.73$) using \citeauthor{Kingsburgh1994}'s \textit{icf} methods \citep[chlorine using][]{Liu2000}. Although the nitrogen composition could be affected by mixing processes prior to the AGB and during the AGB phase, we get [N/H]~=~$-0.63$ (or $-0.95$). The \fcliii\ CELs are typically weak, so their corresponding metallicity may not be accurate. Sulfur abundances in metal-poor stars are expected to behave  the same as metallicity \citep[see][]{Takada-Hidai2005,Spite2011}. However, the \fsii\ lines could be affected by the shock-ionization \citep[see e.g.][]{Danehkar2018a} when dense, small-scale structures propagate through the previously ejected material \citep{Dopita2017,Ali2017}. It is also well known that sulfur abundance derived from optical emission lines in PNe is typically lower than the expected metallicity, which is known as the sulfur anomaly in PNe \citep{Henry2012}. We also found [Ne/H]~=~$-0.30$  (or $-0.82$). AGB nucleosynthesis may slightly increase the neon abundance \citep{Karakas2009}. Thus, the metallicities deduced from argon and oxygen could be more accurate than those derived from sulfur. chlorine, and neon. Interestingly, we notice that the [O/H] and [Ar/H] values are roughly the same, so the metallicity of IC\,4997 is associated with [X/H]~$\lesssim$~$-0.75$. The CEL analysis implies that this PN likely evolved from a metal-poor progenitor. However, we found that [O/H]~=~$0.15$ from the ORL abundance analysis, which is a little higher than the solar composition.

Our CEL and ORL elemental results correspond to $Z\approx 0.004$ and $Z \approx 0.012$, respectively. The CEL elemental abundances derived with the first physical conditions $N($O$)/N($H$)=8.7 \times 10^{-5}$, $N($N$)/N($H$)=1.6\times 10^{-5}$ and $N($Ne$)/N($H$)=4.2\times 10^{-5}$ (by number) are roughly close to the AGB yield predictions of the progenitor mass between 2.25 and 2.5$M_{\odot}$ and the metallicity between $Z=0.004$ and $0.008$ \citep{Karakas2007}. Furthermore, the CEL chemical abundances by mass, $X($O$)/X($He$)=3.3\times 10^{-3}$ (or $1.5\times 10^{-3}$), $X($N$)/X($He$)=5.3\times 10^{-4}$  (or $2.51\times 10^{-4}$) and $X($Ne$)/X($He$)=2\times 10^{-3}$ (or $6.1\times 10^{-4}$) are consistent with the wind predictions of the AGB nucleosynthesis models with the progenitor mass within 1.9--2.25$M_{\odot}$ and the metallicity $0.004 < Z< 0.008$ \citep{Karakas2010}. 

\section{Stellar Characteristics}
\label{ic4997:stellar}

Table~\ref{tab:result:stellar} lists the ionizing luminosities ($L_{\rm UV}$) and effective temperatures ($T_{\rm eff}$) of the central star  derived from the UV \textit{IUE} archival data and the \foiii\ nebular fluxes for different epochs during the 13 years from 1978 to 1991. We modeled the UV continuum of each observing program using polynomial functions fitted to the combined SWP and LWR \textit{IUE} spectra. We dereddened the fitted continuum using $c({\rm H}\beta) = 0.645$ and the UV standard extinction function of \citet{Seaton1979a} with $R_V = 3.1$, and then obtained the integrated flux $F_{\rm UV}$ over the wavelength range 1250--3000\,\AA. The ionizing UV luminosity is then estimated as $L_{\rm UV} = 4 \pi D^2 F_{\rm UV}$, where the distance is taken to be $D=4.32_{-2.17}^{+1.90}$ kpc \citep{Bailer-Jones2018}. The effective temperature $T_{\rm eff}$ of the central star in each year was also estimated from the \foiii\ $\lambda$4959 flux measurement by \citet{Kostyakova2009} using the empirical formulas of \citet{Dopita1990,Dopita1991}. 

\begin{figure*}
\begin{center}
\includegraphics[width=3.4in, trim = 0 0 0 0, clip, angle=0]{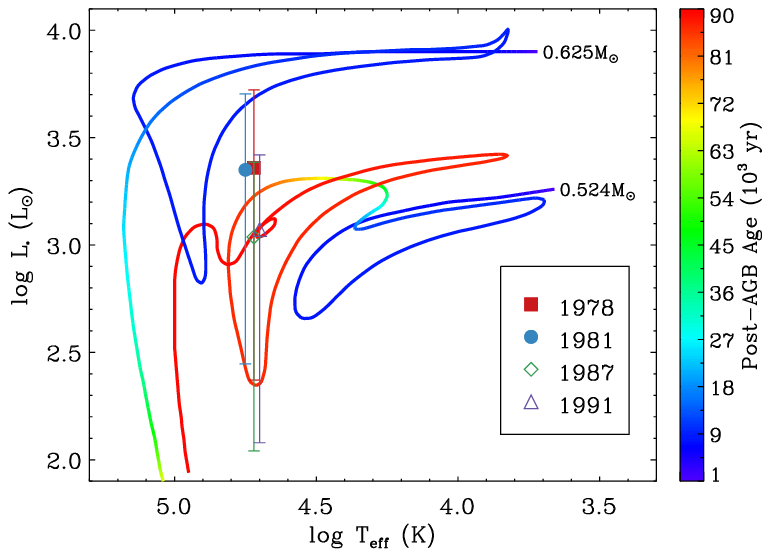}%
\includegraphics[width=3.4in, trim = 0 0 0 0, clip, angle=0]{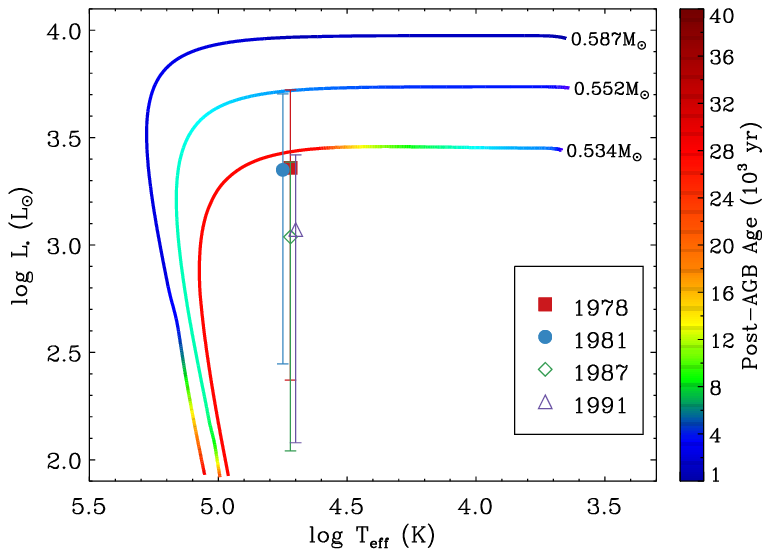}
\end{center}
\caption{The stellar parameters of the central star of IC\,4997 ($T_{\rm eff}$ and $L_{\rm UV}$ listed in Table~\ref{tab:result:stellar}) in four different years (1978, 1981, 1987, and 1991) on the Hertzsprung--Russell diagram of the LTP/VLTP helium-burning models (left panel) with $(M_{\rm init}, M_{\rm \star}) = (1{\rm M}_{\odot}, 0.524{\rm M}_{\odot})$ and $(3{\rm M}_{\odot}, 0.625{\rm M}_{\odot})$ calculated by \citet{Bloecker1995}, and  of the hydrogen-burning models (right panel) with $(M_{\rm init}, M_{\rm \star}) = (0.9{\rm M}_{\odot}, 0.534{\rm M}_{\odot})$, $(1{\rm M}_{\odot}, 0.552{\rm M}_{\odot})$, and $(1.75{\rm M}_{\odot}, 0.587{\rm M}_{\odot})$ generated by \citet{MillerBertolami2016}. The color bar displays the post-AGB ages (unit in $10^3$ years). 
\label{ic4997:hr:diagram}%
}%
\end{figure*}

Figure~\ref{ic4997:hr:diagram} shows the evolutionary tracks of the LTP/VLTP hydrogen-burning models calculated by \citet{Bloecker1995} for the progenitor stars with zero-age main sequence masses of $M_{\rm init}=1{\rm M}_{\odot}$ and $3{\rm M}_{\odot}$ evolving to the degenerate cores with stellar masses of $M_{\star}=0.524{\rm M}_{\odot}$ and $0.625{\rm M}_{\odot}$, respectively (left panel); as well as the evolutionary tracks of the hydrogen-burning models computed by \citet{MillerBertolami2016} for the progenitor masses of $M_{\rm init}=0.9{\rm M}_{\odot}$, $1{\rm M}_{\odot}$, and $1.75{\rm M}_{\odot}$ leading to the degenerate stars with final masses of $M_{\star}=0.534{\rm M}_{\odot}$, $0.552{\rm M}_{\odot}$, and $0.587{\rm M}_{\odot}$ (right panel). Typically, the bolometric luminosity of a hot degenerate core ($T_{\rm eff}\gtrsim 40$\,kK) is mostly from the UV ionizing luminosity, so $L_{\rm UV}$ is roughly associated with the stellar luminosity. In Figure~\ref{ic4997:hr:diagram}, we plotted the locations of $T_{\rm eff}$ and $L_{\rm UV}$ listed in Table~\ref{tab:result:stellar} on the Hertzsprung--Russell diagram. It can be seen that the stellar parameters in four different epochs (1978, 1981, 1987, and 1991) are between the progenitor stars with $M_{\rm init}=1{\rm M}_{\odot}$ and $3{\rm M}_{\odot}$, so the initial mass is close to $2{\rm M}_{\odot}$ as suggested by the AGB models  in \S\,\ref{ic4997:abund}. We notice that the PN age of $t_{\rm age} = 860_{-270}^{+300}$ years in \S\,\ref{ic4997:results:flux} is in agreement with the post-AGB ages of the evolutionary track of the helium-burning model with the progenitor mass of $M_{\rm init}=3{\rm M}_{\odot}$, whereas the nebula must be much older in order to agree with the helium-burning model with the progenitor mass of $M_{\rm init}=1{\rm M}_{\odot}$ and the hydrogen-burning model with $M_{\rm init}=0.9{\rm M}_{\odot}$. 

\begin{table}
\caption{Stellar parameters from 1978 to 1991. \label{tab:result:stellar}}
\footnotesize
\begin{center}
\begin{tabular}{lcccc}
\noalign{\smallskip}
\hline
\noalign{\smallskip}
 Year  &  EC & $T_{\rm eff}$ (K) & $F_{\rm UV}$ ($\frac{\rm erg}{{\rm s}\,{\rm cm}^{2}}$) & $L_{\rm UV}$ (L$_{\odot}$) \\
\noalign{\smallskip}
\hline
\noalign{\smallskip}
1978  &  2.2  &  53,000  &  $3.92 \times 10^{-9}$   &  $2285_{-2050}^{+2990}$  \\
\noalign{\smallskip}
1979  &  2.3  &  55,100  &  -- & -- \\
\noalign{\smallskip}
1980  &  --   & --       &  $4.04 \times 10^{-9}$   &  $2360_{-2120}^{+2930}$  \\
\noalign{\smallskip}
1981  &  2.4  & 55,600   &   $3.84 \times 10^{-9}$   &  $2240_{-1960}^{+2820}$  \\
\noalign{\smallskip}
1986  &  1.6  & 45,800   &  -- & -- \\
\noalign{\smallskip}
1987  &  2.1  & 52,500   &  $1.87 \times 10^{-9}$   &  $1090_{-980}^{+1340}$   \\
\noalign{\smallskip}
1989  & 1.8   & 48,000   &  -- & -- \\
\noalign{\smallskip}
1991  & 2.0   & 50,400   &   $2.02 \times 10^{-9}$   &  $1180_{-1060}^{+1450}$  \\
\noalign{\smallskip}
\hline
\end{tabular}
\end{center}
\begin{tablenotes}
\item[1]\textbf{Note.} The excitation class (EC) derived from the \foiii\ $\lambda$4959 flux measurements using the formula of \citet{Dopita1990}, the effective temperature ($T_{\rm eff}$) deduced from the EC using the empirical relation \citep{Dopita1991}. The integrated flux ($F_{\rm UV}$) derived from the combined SWP and LWR \textit{IUE} spectra dereddened with $c({\rm H}\beta) = 0.645$ using the extinction function of \citet{Seaton1979a}. The ionizing UV luminosity ($L_{\rm UV}$) estimated at $D=4.32_{-2.17}^{+1.90}$ kpc \citep{Bailer-Jones2018}.
\end{tablenotes}
\end{table}

\section{Nebular Variability}
\label{ic4997:variability}

\begin{figure}
\begin{center}
\includegraphics[width=3.3in, trim = 0 25 10 20, clip, angle=0]{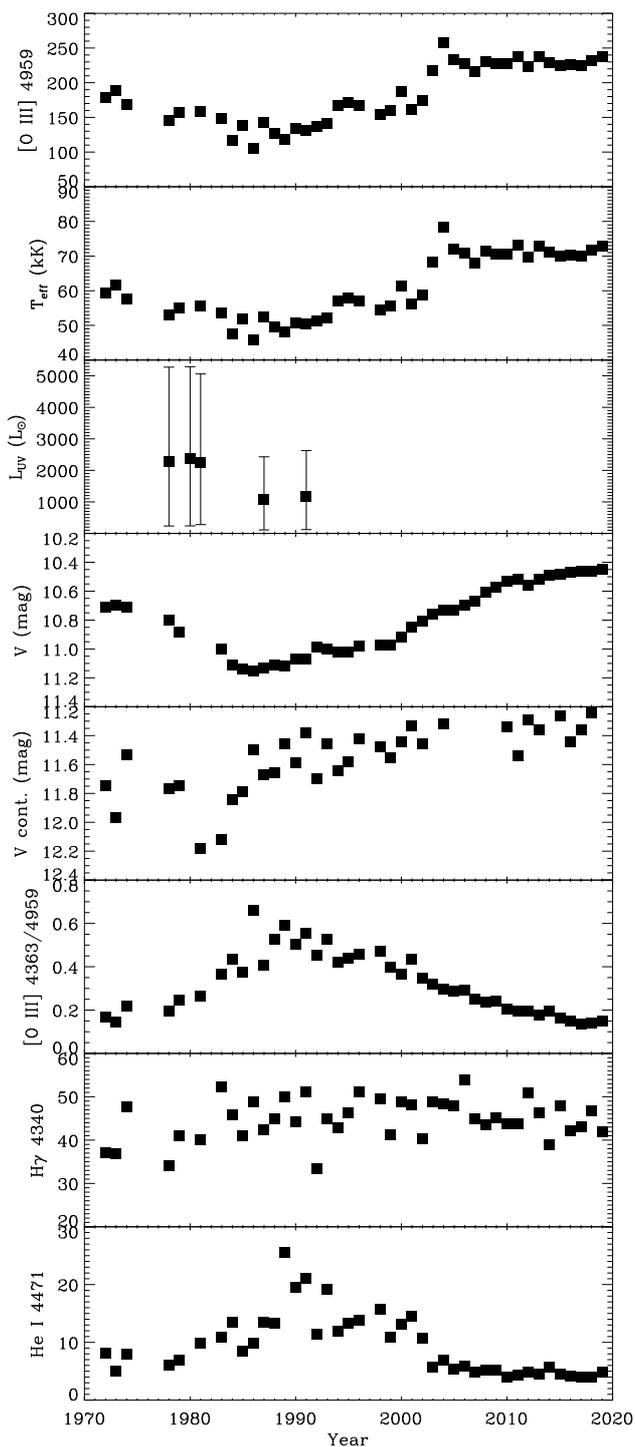}
\end{center}
\caption{Variation of the PN IC\,4997 over the period 1972--2019. From top to bottom, the \foiii\ $\lambda$4959 observed flux, the effective temperature $T_{\rm eff}$(kK) of the central star based on  the \foiii\ $\lambda$4959 flux, 
the UV luminosity  $L_{\rm UV}$(L$_{\odot}$) at $D=4.32_{-2.17}^{+1.90}$ kpc  from the IUE observations listed in Table~\ref{tab:result:stellar}, 
the total $V$-magnitude, the continuum (stellar+nebular) $V$-magnitude estimated by removing emission line contributions to the V-band, the \foiii\ $\lambda$4363/$\lambda$4959 flux ratio, the H$\gamma$ $\lambda$4340 observed flux, and the \hei\ $\lambda$4471 observed flux (where H$\beta=100$). 
The plot was compiled using the 1972--2002 data from \citet{Kostyakova2009}, 2003--2009 from \citet{Burlak2010}, 2014 from the present work, 
and 2010--2019 from \citet{Arkhipova2020}. 
\label{ic4997:diagnostic:variability}%
}%
\end{figure}

The PN IC 4997 is well known for its variable brightness and emission lines. Figure~\ref{ic4997:diagnostic:variability} shows the variability of different spectrum elements over the period 1972--2019, which includes 1972--2002 from \citet{Kostyakova2009}, 2003--2009 from \citet{Burlak2010}, 2010--2019 from \citet{Arkhipova2020}, and the present study (2014): the \foiii\ $\lambda$4959 flux, the stellar temperature ($T_{\rm eff}$), 
the UV luminosity  $L_{\rm UV}$ (listed in Table~\ref{tab:result:stellar}), 
the total integral $V$-magnitude, the integral continuum $V$-magnitude, the \foiii\ $\lambda$4363/$\lambda$4959 flux ratio, 
the H$\gamma$ $\lambda$4340 flux, and the \hei\ $\lambda$4471 flux.  
We estimated the integral continuum $V$-magnitude by removing the total emission line fluxes presented in the $V$-band using the available $V$-magnitudes and the line flux measurements from \citet{Kostyakova2009} and \citet{Arkhipova2020} ($V$-mag for 1967--2007 provided by N.\,P. Ikonnikova), and assuming the  Johnson-Cousins standard $V$ bandpass \citep{Bessell2005}. The integral continuum $V$-magnitude contains both the stellar and nebular continua. As the stellar temperature is relativity high ($\gtrsim 40$\,kK), we expect to have a significant nebular contribution to the  $V$ magnitude, so we cannot use it to calculate the absolute bolometric magnitude of the central star.

It is known that the \foiii\ $\lambda\lambda$4959,5007 doublet is an indicator of the effective temperature ($T_{\rm eff}$) of the central star \citep{Dopita1990,Dopita1991,Reid2010}. As the \foiii\ $\lambda$5007 emission line might be saturated, we estimated $T_{\rm eff}$ from the \foiii\ $\lambda$4959 emission line. We obtained the excitation class using ${\rm EC} = 1.485 \times I($\foiii$\lambda 4959)/I({\rm H}\beta)$. We then estimated the stellar effective temperature from the excitation class using the empirical relation of \citet{Dopita1991}. The continuum $V$ magnitude might be an indicator of the stellar luminosity, but it may be contaminated by the nebular continuum. As seen in Fig.~\ref{ic4997:diagnostic:variability}, the stellar temperature slightly decreased from 62,000\,K in 1973 to 46\,K in 1986, but gradually increased to 78\,K by 2004. It dropped to 70\,K in 2005 and roughly remained the same until 2014. It is seen in Fig.~\ref{ic4997:diagnostic:variability} that the UV luminosity is lower in 1987--1991 compared to 1978--1981. The continuum $V$ magnitude also decreased to its lowest brightness in 1981-1883 but was then slightly increased after 1984, and experienced some instant increases and decreases over 1986--2005 and 2010--2019, which might be associated with multiple outbursts. The 1980--81 UV luminosity was brighter when the stellar temperature was around 55,000\,K, whereas it was fainter in 1987--1991 after the stellar temperature reached its lowest value of 45,800\,K in 1986. As seen in Figure~\ref{ic4997:hr:diagram}, these changes in the stellar luminosity and the stellar temperature over the period 1978--1991 are consistent with the LTP/VLTP evolutionary tracks of the helium-burning models of \citet{Bloecker1995}. The figure also shows the hydrogen-burning evolutionary tracks from \citet{MillerBertolami2016}, which do not seem to provide an acceptable match, apart from the model with a progenitor mass of $0.9{\rm M}_{\odot}$, whose post-AGB age does not agree with the nebular age. We also notice a sharp increase in the \hei\ $\lambda$4471 flux in 1991, which was gradually decreasing during 1992--1998, and some further declines until 2003. In our case, the star was likely returning from the PN stage before 1980 to the AGB phase in 1985--1995, and then gradually returned to the PN phase after 1995.

Figure~\ref{ic4997:diagnostic:variability} also depicts the variation of the \foiii\ $\lambda$4363/$\lambda$4959 flux ratio, which is sensitive to the electron temperature and density. If the electron density was known and below the critical density of the density-diagnostic lines, we could accurately derive the electron temperature $T_{\rm e}$ from this flux ratio. The nebula temperature may have been very low in 1972, but gradually increased in 1981. Over the period 1983--2002, it was constantly very high, while it was gradually decreasing from 2003 to 2019. In particular, the \hei\ $\lambda$4471 emission line approximately follows the pattern of the \foiii\ $\lambda$4363/$\lambda$4959 flux ratio. This could be related to an increase in the electron temperature or/and helium elemental abundance. A sudden increase in the helium abundance might be related to a helium-shell flash, which should not typically occur in the post-AGB phase. However, a (very) late helium-shell flash happens in the VLTP/LTP born-again scenario \citep{Werner2006}, which results in H-deficient post-AGB stars and can also eject metal-rich material into the previously expelled H-rich nebula \citep{Liu2003,Liu2004b}. 

The observed Balmer $\lambda$4340 emission line has an average value of 45, and slightly varies between 33 and 54 during 1972--2019. We note that some parts of the nebula are heavily reddened \citep[][Fig.\,1]{Miranda1996}, so the variation in the Balmer $\lambda$4340 emission line can be explained by the ordinations and locations of different slits used in the 1972--2019 observations. The \foiii\ $\lambda$5007 (or $\lambda$4959) emission line flux (relative to H$\beta$) has typically little variation over the nebula \citep[see e.g.][]{Danehkar2018a}, so the large variation in the \foiii\ line is most likely associated with thermal effects in different epochs. As this PN has the excitation class of ${\rm EC}=2.5^{+1.3}_{-0.9}$ 
\citep[using the formula from][]{Dopita1990} and no strong \heii\ line, the \hei\ emission line should not have large variability. The large variation in the \hei\ emission line is most likely attributed to the physical conditions ($T_{\rm e}$ and $N_{\rm e}$) or/and the helium abundance because of an LTP/VLTP event.

IC 4997 has a high galactic latitude ($b = -10.98^{\circ}$) as well as a high systemic velocity \citep[$v_{\rm hel}=-66.2$\,km\,s$^{-1}$;][]{Durand1998}. Its low metallicity ($Z \approx 0.004$), high galactic latitude, and high systemic velocity indicate that its initial main sequence mass may be about twice the solar mass. The progenitor may be an old disk population II star. In order to further understand its evolution, it is important to derive light and heavy $s$-process elemental abundances from high resolution spectra since LTP and VLTP objects are typically expected to have overabundant $s$-process elements \citep{Lawlor2003}.

\section{Conclusion}
\label{ic4997:summary}

We have determined the physical and chemical properties of the PN IC 4997 using the NOT/FIES observations collected in 2014. The CELs and ORLs have been employed to conduct plasma diagnostics. 
The temperatures derived from the \foi, \fnii, and \fariii\ line ratios are $\sim$ 12\,900, 18\,000, and 14\,300\,K, respectively. Moreover, the \foiii\ temperature could be $\lesssim$ 24\,000\,K at densities $\gtrsim  10^5$\,cm$^{-3}$. A least-squares minimization method on the \oii\ and \nii\ ORLs also pointed to  temperatures of $7400^{+17400}_{-5300}$ and $6700^{+8100}_{-2900}$\,K, respectively, which are lower than those from CELs.  The \fcliii\ and \fariv\ line ratios also imply relatively high densities of $\approx 3.3 \times 10^4$ and $2.97 \times 10^5$\,cm$^{-3}$, respectively. We should also note large CEL--ORL temperature dichotomies of $7410_{-1380}^{+1470}$ and $7210_{-7210}^{+4500}$\,K deduced from low-excitation and high-excitation lines, respectively. 

We have obtained the ionic abundances and total abundances of different elements using two different physical conditions and two different sets of \textit{icf} formulas (see Table~\ref{tab:result:total:abund}). The elemental abundances of oxygen and argon derived from the CELs using \citet{Kingsburgh1994}'s \textit{icf} methods suggest that the PN was expelled from a metal-poor progenitor with [X/H]~$\approx$~$-0.75$ (or $-1.1$ see Table~\ref{tab:result:total:abund}). However, the oxygen elemental abundance deduced from the \oii\ ORLs with \citet{Wang2007}'s \textit{icf} formula is associated with a metallicity of [O/H]~$\approx$~$0.15$. The abundances of the ions N$^{2+}$ and O$^{2+}$ derived from the ORLs are greater than those calculated from the CELs, corresponding to relatively large ADFs: ADF(N$^{2+})\gtrsim 7$ and ADF(O$^{2+})\gtrsim 8$. The bi-abundance model, which has been proposed to solve the abundance discrepancy problem \citep{Liu2003,Liu2004b}, indicates that the heavy element ORLs are mostly emitted from cool metal-rich small structures. The applicability of the bi-abundance hypothesis has been supported by photoionization models \citep{Ercolano2003a,Yuan2011,Danehkar2018b,Gomez-Llanos2020}. Moreover, some observations suggest that most heavy-element ORLs originate from a cold, dense component within the nebulae \citep[e.g.,][]{Richer2013,Richer2019}. 

The ORL abundances in IC\,4997 may be related to the metal-rich ejecta of a (very-) late thermal pulse event, i.e., the born-again scenario. However, our derived ORL C/O abundance ratio of $\approx 0.35$ (or $0.16$) with $icf$ formulas from \citet{Wang2007} \citep[or][]{Delgado-Inglada2014} is less than unity, which disagrees with the theoretical yields of available born-again models \citep{Herwig2001,Althaus2005,Werner2006}. Previously, the C/O ratios in the H-deficient knots of the born-again PN Abell 30 \citep{Wesson2003} and Abell 58 \citep{Wesson2008} were also found to be less than unity, which is in disagreement with the born-again theoretical predictions. Moreover, about half of the 12 PNe with H-deficient central stars analyzed by \citet{Garcia-Rojas2013} were found to have C/O $< 1$, which likely evolved from progenitor masses of less than $2 M_{\odot}$ according to stellar evolution models. Based on the AGB models \citep{Karakas2010}, IC\,4997 could also be descended from a progenitor star with $\sim 2 M_{\odot}$. 

Alternatively, large abundance discrepancies in PNe could be linked to binary central stars \citep[][]{Corradi2015,Jones2016,Bautista2018,Wesson2018,Jacoby2020,Garcia-Rojas2022}. In particular, \citet{Corradi2015} found ADFs higher than 50 in the PNe Abell\,46 and Ou\,5 surrounding post common-envelope binary stars. Moreover, \citet{Jones2016} derived an ADF of $\sim 20$ in NGC\,6778 around a short-period binary.
More recently, light curve monitoring of the central star of the born-again PN Abell\,30 shows $\sim 1$ day periodicity that is highly indicative of a binary system \citep{Jacoby2020}. 
Thus, the possibility of a binary central star in IC\,4997 should be explored, which can explain its variable brightness and abundance discrepancies. %Accretion onto a companion could also scrape off the hydrogen-rich surface layer, leaving behind a hydrogen-deficient core.
 
In conclusion, the nebular and stellar variability in IC\,4997 as recorded since half a century ago may be an indicator of some outbursts during the PN phase, such as an LTP/VLTP event or the presence of a binary companion. It is possible that the ORL abundances in this PN are mostly associated with metal-rich knots ejected by recent outbursts via either LTP/VLTP or a binary channel. In the next few decades, more in-depth studies of this object will help us better understand how this PN and its central star evolved over time.

\section*{Acknowledgements}

Based on observations made with the Nordic Optical Telescope (NOT), operated by the Nordic Optical Telescope Scientific Association (NOTSA) at the Observatorio del Roque de los Muchachos (La Palma, Spain) of the Instituto de Astrof{\'{\i}}sica de Canarias. We thank J.\,J. D{\'{\i}}az-Luis and D.\,A. Garc{\'{\i}}a-Hern{\'a}ndez for providing us with the NOT/FIES spectra of the PN IC\,4997,  N.\,P. Ikonnikova  and V.\,P. Arkhipova for sharing their V-magnitude measurements of IC\,4997 from 1967 to 2007, and the anonymous referee for valuable remarks. We gratefully thank Sergio Armas for the FIES calibration data and  J. J. D{\'{\i}}az-Luis for the data reduction with FIEStool. This research is based on observations made with the \textit{International Ultraviolet Explorer}, obtained from the MAST data archive at the Space Telescope Science Institute, which is operated by the Association of Universities for Research in Astronomy, Inc., under NASA contract NAS 5-26555.
\linebreak

\noindent\hspace{8pt}\textit{Software:} IDL Astronomy User's Library \citep{Landsman1993}, IDL Coyote Library \citep{Fanning2011}, IDL proEQUIB Library \citep{Danehkar2018}, IDL AtomNeb Library \citep{Danehkar2019}.

\section*{Data Availability}

The observational raw data underlying this article can be retrieved from the NOT data archive website at  \href{http://www.not.iac.es/archive/}{http://www.not.iac.es/archive/}, and the reduced data will be shared on reasonable request to the corresponding author.

%%%%%%%%%%%%%%%%%%%%%%%%%%%%%%%%%%%%%%%%%%%%%%%%%%

%%%%%%%%%%%%%%%%%%%% REFERENCES %%%%%%%%%%%%%%%%%%

% The best way to enter references is to use BibTeX:

%\bibliographystyle{mnras2}
%\bibliography{references}

%\newpage 

%\appendix

%\section*{Supplementary Material}
%\label{ic4997:supplementary}

%\subsection*{Identified Emission Lines}

%%%%%%%%%%%%%%%%%%%%%%%%%%%%%%%%%%%%%%%%%%%%%%%%%%

%\begin{appendix}

%The following table is available for the electronic edition of this article:
%\\~\\
%\textbf{Table A1.} Identified emission lines in the FIES Spectrum of the PN IC\,4997.
%\
%Table \ref{tab:spectrum:lines} is available for this article that presents the identified emission lines in the NOT/FIES observations of  IC\,4997 used in our analysis. Figure \ref{fig:ic4997:spectrum:all} is also available for this article that shows the NOT/FIES combined spectrum of IC\,4997 taken with the exposure time of 1800 seconds, including the Gaussian curves fitted using the IDL library MGFIT.

%\newpage

\appendix

\section*{Supplementary Material}
\label{rtcru:supplementary}

\section{Identified Emission Lines}
\label{IC4997:sec:emission}

Table \ref{tab:spectrum:lines} is available for this article that presents the identified emission lines in the NOT/FIES observations of  IC\,4997 used in our analysis. 
%Figure \ref{fig:ic4997:spectrum:all} is also available for this article that shows the NOT/FIES combined spectrum of IC\,4997 taken with the exposure time of 1800 seconds, including the Gaussian curves fitted using the IDL library MGFIT.

\newpage
\FloatBarrier
\normalsize

\begin{table*}
%\begin{center}
\caption{Identified emission lines in the FIES Spectrum of the PN IC\,4997.
\label{tab:spectrum:lines}
}
%\centering
\scriptsize
\begin{center}
\begin{tabular}{lllccccccccc}
\noalign{\smallskip}
\hline
\noalign{\smallskip}
 $\lambda_{\rm lab}$  & Ion  & $\lambda_{\rm obs}$ & $F(\lambda)$ & $\varepsilon_{F(\lambda)}$(\%) & $I(\lambda)$ & $\varepsilon_{I(\lambda)}$(\%) & Mult & Lower term & Upper term & g1 & g2 \\
\noalign{\smallskip}
\hline
\noalign{\smallskip}
   3697.15 & H\,{\sc i}                     &    3696.13 &      0.845 & $\pm  0.6$ &      1.246 & $\pm  1.6$ & H17      & 2p+ 2P*          & 17d+ 2D          & 8        & *        \\
\noalign{\smallskip}
   3711.97 & H\,{\sc i}                     &    3711.04 &      1.084 & $\pm  2.4$ &      1.593 & $\pm  3.5$ & H15      & 2p+ 2P*          & 15d+ 2D          & 8        & *        \\
\noalign{\smallskip}
   3726.03 & [O\,{\sc ii}]                  &    3725.15 &     11.379 & $\pm  1.8$ &     16.664 & $_{ -3.1}^{+  2.8}$ & F1       & 2p3 4S*          & 2p3 2D*          & 4        & 4        \\
\noalign{\smallskip}
   3728.82 & [O\,{\sc ii}]                  &    3727.94 &      3.667 & $\pm  1.8$ &      5.366 & $_{ -3.0}^{+  2.8}$ & F1       & 2p3 4S*          & 2p3 2D*          & 4        & 6        \\
\noalign{\smallskip}
   3734.37 & H\,{\sc i}                     &    3733.33 &      1.776 & $\pm  0.4$ &      2.595 & $\pm  1.5$ & H13      & 2p+ 2P*          & 13d+ 2D          & 8        & *        \\
\noalign{\smallskip}
   3750.15 & H\,{\sc i}                     &    3749.02 &      2.391 & $\pm  0.8$ &      3.480 & $_{ -2.0}^{+  1.7}$ & H12      & 2p+ 2P*          & 12d+ 2D          & 8        & *        \\
\noalign{\smallskip}
   3770.63 & H\,{\sc i}                     &    3769.59 &      2.718 & $\pm  0.3$ &      3.934 & $\pm  1.5$ & H11      & 2p+ 2P*          & 11d+ 2D          & 8        & *        \\
\noalign{\smallskip}
   3791.27 & O\,{\sc iii}                   &    3790.45 &      0.228 & $\pm  1.5$ &      0.328 & $\pm  2.4$ & V2       & 3s 3P*           & 3p 3D            & 5        & 5        \\
\noalign{\smallskip}
   3797.90 & H\,{\sc i}                     &    3796.85 &      3.567 & $\pm  0.3$ &      5.125 & $\pm  1.5$ & H10      & 2p+ 2P*          & 10d+ 2D          & 8        & *        \\
\noalign{\smallskip}
   3819.62 & He\,{\sc i}                    &    3818.56 &      0.892 & $\pm  0.6$ &      1.274 & $\pm  1.7$ & V22      & 2p 3P*           & 6d 3D            & 9        & 15       \\
\noalign{\smallskip}
   3834.89 & He\,{\sc ii}                   &    3834.32 &      0.768 & $\pm  1.6$ &      1.092 & $\pm  2.6$ & 4.18     & 4f+ 2F*          & 18g+ 2G          & 32       & *        \\
\noalign{\smallskip}
   3835.39 & H\,{\sc 9}                     &    3834.33 &      4.872 & $\pm  0.3$ &      6.928 & $\pm  1.5$ & H9       & 2p+ 2P*          & 9d+ 2D           & 8        & *        \\
\noalign{\smallskip}
   3868.75 & [Ne\,{\sc iii}]                &    3867.62 &     92.031 & $\pm  0.6$ &    129.616 & $\pm  1.6$ & F1       & 2p4 3P           & 2p4 1D           & 5        & 5        \\
\noalign{\smallskip}
   3889.65 & N\,{\sc i}                     &    3888.50 &     11.943 & $\pm  0.4$ &     16.718 & $\pm  1.4$ & V52      & 3p  4P*          & 3d  4P           & 2        & 2        \\
\noalign{\smallskip}
   3926.54 & He\,{\sc i}                    &    3925.39 &      0.105 & $\pm  2.9$ &      0.145 & $\pm  4.0$ & V58      & 2p 1P*           & 8d 1D            & 3        & 5        \\
\noalign{\smallskip}
   3967.46 & [Ne\,{\sc iii}]                &    3966.31 &     27.238 & $\pm  1.0$ &     37.243 & $_{ -2.2}^{+  1.8}$ & F1       & 2p4 3P           & 2p4 1D           & 3        & 5        \\
\noalign{\smallskip}
   3970.07 & H\,{\sc 7}                     &    3968.96 &     12.538 & $\pm  0.1$ &     17.129 & $\pm  1.2$ & H7       & 2p+ 2P*          & 7d+ 2D           & 8        & 98       \\
\noalign{\smallskip}
   4009.26 & He\,{\sc i}                    &    4008.17 &      0.158 & $\pm  0.9$ &      0.213 & $\pm  1.6$ & V55      & 2p 1P*           & 7d 1D            & 3        & 5        \\
\noalign{\smallskip}
   4068.60 & [S\,{\sc ii}]                  &    4067.50 &      1.198 & $\pm  1.4$ &      1.586 & $\pm  2.1$ & F1       & 2p3 4S*          & 2p3 2P*          & 4        & 4        \\
\noalign{\smallskip}
   4076.35 & [S\,{\sc ii}]                  &    4075.68 &      0.403 & $\pm  2.4$ &      0.532 & $\pm  3.3$ & F1       & 2p3 4S*          & 2p3 2P*          & 2        & 4        \\
\noalign{\smallskip}
   4101.74 & H\,{\sc 6}                     &    4100.61 &     20.347 & $\pm  0.6$ &     26.652 & $\pm  1.2$ & H6       & 2p+ 2P*          & 6d+ 2D           & 8        & 72       \\
\noalign{\smallskip}
   4110.78 & O\,{\sc ii}                    &    4109.51 &      0.011 & $\pm  5.2$ &      0.014 & $\pm  6.8$ & V20      & 3p 4P*           & 3d 4D            & 4        & 2        \\
\noalign{\smallskip}
   4119.22 & O\,{\sc ii}                    &    4118.08 &      0.047 & $\pm  2.0$ &      0.061 & $\pm  2.7$ & V20      & 3p 4P*           & 3d 4D            & 6        & 8        \\
\noalign{\smallskip}
   4120.84 & He\,{\sc i}                    &    4119.74 &      0.028 & $\pm  3.3$ &      0.036 & $\pm  4.2$ & V16      & 2p 3P*           & 5s 3S            & 9        & 3        \\
\noalign{\smallskip}
   4143.76 & He\,{\sc i}                    &    4142.51 &      0.167 & $\pm  0.9$ &      0.216 & $\pm  1.4$ & V53      & 2p 1P*           & 6d 1D            & 3        & 5        \\
\noalign{\smallskip}
   4153.30 & O\,{\sc ii}                    &    4151.94 &      0.047 & $\pm  4.2$ &      0.061 & $\pm  5.4$ & V19      & 3p 4P*           & 3d 4P            & 4        & 6        \\
\noalign{\smallskip}
   4253.86 & O\,{\sc ii}                    &    4252.76 &      0.074 & $\pm  2.4$ &      0.092 & $\pm  3.1$ & V101     & 3d 2G            & 4f 2[5]*         & 10       & 10       \\
\noalign{\smallskip}
   4317.14 & O\,{\sc ii}                    &    4315.90 &      0.073 & $\pm  1.2$ &      0.089 & $\pm  1.6$ & V2       & 3s 4P            & 3p 4P*           & 2        & 4        \\
\noalign{\smallskip}
   4319.63 & O\,{\sc ii}                    &    4318.45 &      0.034 & $\pm  5.3$ &      0.041 & $\pm  6.6$ & V2       & 3s 4P            & 3p 4P*           & 4        & 6        \\
\noalign{\smallskip}
   4340.47 & H\,{\sc 5}                     &    4339.41 &     39.021 & $\pm  0.8$ &     47.109 & $\pm  1.1$ & H5       & 2p+ 2P*          & 5d+ 2D           & 8        & 50       \\
\noalign{\smallskip}
   4349.43 & O\,{\sc ii}                    &    4348.24 &      0.069 & $\pm  2.6$ &      0.083 & $\pm  3.2$ & V2       & 3s 4P            & 3p 4P*           & 6        & 6        \\
\noalign{\smallskip}
   4358.81 & [Fe\,{\sc ii}]                 &    4358.23 &      0.015 & $\pm 13.4$ &      0.018 & $\pm 16.1$ & F7       & 3d6 3D           & 3d6 3P1          & 2        & 4        \\
\noalign{\smallskip}
   4363.21 & [O\,{\sc iii}]                 &    4361.87 &     45.308 & $\pm  1.0$ &     54.263 & $\pm  1.3$ & F2       & 2p2 1D           & 2p2 1S           & 5        & 1        \\
\noalign{\smallskip}
   4387.93 & He\,{\sc i}                    &    4386.71 &      0.600 & $\pm  0.9$ &      0.712 & $\pm  1.2$ & V51      & 2p 1P*           & 5d 1D            & 3        & 5        \\
\noalign{\smallskip}
   4416.97 & O\,{\sc ii}                    &    4415.70 &      0.028 & $\pm  6.7$ &      0.033 & $_{ -8.3}^{+  7.8}$ & V5       & 3s 2P            & 3p 2D*           & 2        & 4        \\
\noalign{\smallskip}
   4430.94 & Ne\,{\sc ii}                   &    4429.47 &      0.049 & $\pm  3.2$ &      0.057 & $_{ -4.0}^{+  3.8}$ & V61a     & 3d 2D            & 4f 2[4]*         & 6        & 8        \\
\noalign{\smallskip}
   4452.37 & O\,{\sc ii}                    &    4451.00 &      0.023 & $\pm  3.9$ &      0.027 & $\pm  4.6$ & V5       & 3s 2P            & 3p 2D*           & 4        & 4        \\
\noalign{\smallskip}
   4471.50 & He\,{\sc i}                    &    4470.26 &      5.327 & $\pm  0.3$ &      6.138 & $\pm  0.6$ & V14      & 2p 3P*           & 4d 3D            & 9        & 15       \\
\noalign{\smallskip}
   4510.91 & N\,{\sc iii}                   &    4509.76 &      0.009 & $\pm 11.9$ &      0.010 & $\pm 14.1$ & V3       & 3s' 4P*          & 3p' 4D           & 2        & 4        \\
\noalign{\smallskip}
   4514.86 & N\,{\sc iii}                   &    4513.02 &      0.013 & $\pm  5.3$ &      0.015 & $\pm  6.5$ & V3       & 3s' 4P*          & 3p' 4D           & 6        & 8        \\
\noalign{\smallskip}
   4530.41 & N\,{\sc ii}                    &    4528.69 &      0.043 & $\pm  2.7$ &      0.049 & $\pm  3.3$ & V58b     & 3d 1F*           & 4f 2[5]          & 7        & 9        \\
\noalign{\smallskip}
   4530.86 & N\,{\sc iii}                   &    4528.99 &      0.165 & $\pm  1.0$ &      0.186 & $\pm  1.2$ & V3       & 3s' 4P*          & 3p' 4D           & 4        & 2        \\
\noalign{\smallskip}
   4562.60 & Mg\,{\sc i}]                   &    4561.02 &      0.153 & $\pm  1.6$ &      0.171 & $_{ -2.1}^{+  1.9}$ &          & 3s2 1S           & 3s3p 3P*         & 1        & 5        \\
\noalign{\smallskip}
   4571.10 & Mg\,{\sc i}]                   &    4569.87 &      0.083 & $\pm  4.2$ &      0.092 & $\pm  5.0$ &          & 3s2 1S           & 3s3p 3P*         & 1        & 3        \\
\noalign{\smallskip}
   4590.97 & O\,{\sc ii}                    &    4589.66 &      0.034 & $\pm  2.5$ &      0.038 & $\pm  3.1$ & V15      & 3s' 2D           & 3p' 2F*          & 6        & 8        \\
\noalign{\smallskip}
   4596.18 & O\,{\sc ii}                    &    4597.75 &      0.132 & $\pm  1.2$ &      0.145 & $\pm  1.4$ & V15      & 3s' 2D           & 3p' 2F*          & 4        & 6        \\
\noalign{\smallskip}
   4609.44 & O\,{\sc ii}                    &    4608.16 &      0.025 & $\pm  2.5$ &      0.027 & $_{ -3.3}^{+  2.9}$ & V92a     & 3d 2D            & 4f F4*           & 6        & 8        \\
\noalign{\smallskip}
   4610.20 & O\,{\sc ii}                    &    4608.92 &      0.012 & $\pm  5.2$ &      0.013 & $_{ -6.7}^{+  6.0}$ & V92c     & 3d 2D            & 4f F2*           & 4        & 6        \\
\noalign{\smallskip}
   4630.54 & N\,{\sc ii}                    &    4632.90 &      0.036 & $\pm  1.4$ &      0.039 & $\pm  1.6$ & V5       & 3s 3P*           & 3p 3P            & 5        & 5        \\
\noalign{\smallskip}
   4634.14 & N\,{\sc iii}                   &    4635.17 &      0.106 & $\pm  2.9$ &      0.115 & $\pm  3.4$ & V2       & 3p 2P*           & 3d 2D            & 2        & 4        \\
\noalign{\smallskip}
   4640.64 & N\,{\sc iii}                   &    4638.94 &      0.140 & $\pm  4.3$ &      0.152 & $_{ -5.5}^{+  4.8}$ & V2       & 3p 2P*           & 3d 2D            & 4        & 6        \\
\noalign{\smallskip}
   4641.81 & O\,{\sc ii}                    &    4639.32 &      0.183 & $\pm  2.5$ &      0.198 & $_{ -3.1}^{+  2.8}$ & V1       & 3s 4P            & 3p 4D*           & 4        & 6        \\
\noalign{\smallskip}
\hline
\end{tabular}
\end{center}
\end{table*}
\setcounter{table}{1}
\begin{table*}
%\caption{-- \textit{continued}}
\contcaption{}
\scriptsize
%\contcaption{}
%\centering
\begin{center}
\begin{tabular}{lllccccccccc}
\noalign{\smallskip}
\hline
\noalign{\smallskip}
 $\lambda_{\rm lab}$  & Ion  & $\lambda_{\rm obs}$ & $F(\lambda)$ & $\varepsilon_{F(\lambda)}$(\%) & $I(\lambda)$ & $\varepsilon_{I(\lambda)}$(\%) & Mult & Lower term & Upper term & g1 & g2 \\
\noalign{\smallskip}
\hline
\noalign{\smallskip}
   4647.42 & C\,{\sc iii}                   &    4646.13 &      0.100 & $\pm  2.0$ &      0.108 & $\pm  2.3$ & V1       & 3s 3S            & 3p 3P*           & 3        & 5        \\
\noalign{\smallskip}
   4649.13 & O\,{\sc ii}                    &    4647.84 &      0.201 & $\pm  1.7$ &      0.217 & $_{ -2.2}^{+  1.9}$ & V1       & 3s 4P            & 3p 4D*           & 6        & 8        \\
\noalign{\smallskip}
   4650.25 & C\,{\sc iii}                   &    4648.96 &      0.057 & $\pm  6.0$ &      0.062 & $_{ -7.3}^{+  6.9}$ & V1       & 3s 3S            & 3p 3P*           & 3        & 3        \\
\noalign{\smallskip}
   4650.84 & O\,{\sc ii}                    &    4647.77 &      0.052 & $\pm  5.1$ &      0.056 & $_{ -6.6}^{+  5.9}$ & V1       & 3s 4P            & 3p 4D*           & 2        & 2        \\
\noalign{\smallskip}
   4658.64 & C\,{\sc iv}                    &    4656.92 &      0.303 & $\pm  1.7$ &      0.326 & $\pm  2.0$ & V8       & 5f 2F*           & 6g 2G            & 14       & 18       \\
\noalign{\smallskip}
   4661.63 & O\,{\sc ii}                    &    4660.53 &      0.110 & $\pm  1.1$ &      0.118 & $\pm  1.3$ & V1       & 3s 4P            & 3p 4D*           & 4        & 4        \\
\noalign{\smallskip}
   4678.14 & N\,{\sc ii}                    &    4677.42 &      0.142 & $\pm  0.6$ &      0.152 & $\pm  0.7$ & V61b     & 3d 1P*           & 4f 2[2]          & 3        & 5        \\
\noalign{\smallskip}
   4701.62 & [Fe\,{\sc iii}]                &    4700.31 &      0.156 & $\pm  1.5$ &      0.165 & $\pm  1.7$ & F3       & 3d6 5D           & 3d6 3F2          & 7        & 7        \\
\noalign{\smallskip}
   4711.37 & [Ar\,{\sc iv}]                 &    4710.21 &      0.103 & $\pm  2.1$ &      0.109 & $\pm  2.4$ & F1       & 3p3 4S*          & 3p3 2D*          & 4        & 6        \\
\noalign{\smallskip}
   4733.91 & [Fe\,{\sc iii}]                &    4732.70 &      0.070 & $\pm  3.5$ &      0.073 & $\pm  4.0$ & F3       & 3d6 5D           & 3d6 3F2          & 5        & 5        \\
\noalign{\smallskip}
   4740.17 & [Ar\,{\sc iv}]                 &    4738.89 &      0.662 & $\pm  0.7$ &      0.692 & $\pm  0.8$ & F1       & 3p3 4S*          & 3p3 2D*          & 4        & 4        \\
\noalign{\smallskip}
   4769.40 & [Fe\,{\sc iii}]                &    4768.23 &      0.076 & $\pm  2.2$ &      0.079 & $\pm  2.5$ & F3       &                  &                  &          &          \\
\noalign{\smallskip}
   4676.24 & O\,{\sc ii}                    &   23381.20 &      0.034 & $\pm  7.5$ &      0.036 & $\pm  9.0$ & V1       & 3s 4P            & 3p 4D*           & 6        & 6        \\
\noalign{\smallskip}
   4861.33 & H\,{\sc 4}                     &    4860.01 &    100.000 & $\pm  0.2$ &    100.002 & $\pm  0.2$ & H4       & 2p+ 2P*          & 4d+ 2D           & 8        & 32       \\
\noalign{\smallskip}
   4958.91 & [O\,{\sc iii}]                 &    4957.72 &    229.307 & $\pm  0.7$ &    221.290 & $\pm  0.9$ & F1       & 2p2 3P           & 2p2 1D           & 3        & 5        \\
\noalign{\smallskip}
   5047.74 & He\,{\sc i}                    &    5046.49 &      0.324 & $\pm  4.8$ &      0.303 & $\pm  5.8$ & V47      & 2p 1P*           & 4s 1S            & 3        & 1        \\
\noalign{\smallskip}
   5191.82 & [Ar\,{\sc iii}]                &    5187.55 &      0.149 & $\pm  1.4$ &      0.132 & $\pm  1.9$ & F3       & 2p4 1D           & 2p4 1S           & 5        & 1        \\
\noalign{\smallskip}
   5270.40 & [Fe\,{\sc iii}]                &    5269.14 &      0.172 & $\pm  2.8$ &      0.148 & $\pm  3.6$ & F1       & 3d6 5D           & 3d6 3P2          & 7        & 5        \\
\noalign{\smallskip}
   5411.52 & He\,{\sc ii}                   &    5410.78 &      0.023 & $\pm  4.0$ &      0.019 & $_{ -4.9}^{+  5.1}$ & 4.7      & 4f+ 2F*          & 7g+ 2G           & 32       & 98       \\
\noalign{\smallskip}
   5517.66 & [Cl\,{\sc iii}]                &    5516.29 &      0.078 & $\pm  1.3$ &      0.062 & $_{ -1.8}^{+  2.0}$ & F1       & 2p3 4S*          & 2p3 2D*          & 4        & 6        \\
\noalign{\smallskip}
   5537.60 & [Cl\,{\sc iii}]                &    5536.38 &      0.239 & $\pm  0.9$ &      0.189 & $\pm  1.6$ & F1       & 2p3 4S*          & 2p3 2D*          & 4        & 4        \\
\noalign{\smallskip}
   5577.34 & [O\,{\sc i}]                   &    5575.80 &      0.117 & $\pm  2.3$ &      0.092 & $_{ -3.0}^{+  3.2}$ & F3       & 2p4 1D           & 2p4 1S           & 5        & 1        \\
\noalign{\smallskip}
   5666.63 & N\,{\sc ii}                    &    5665.05 &      0.020 & $\pm  1.8$ &      0.015 & $\pm  2.6$ & V3       & 3s 3P*           & 3p 3D            & 3        & 5        \\
\noalign{\smallskip}
   5676.02 & N\,{\sc ii}                    &    5674.71 &      0.006 & $\pm  4.7$ &      0.005 & $\pm  5.7$ & V3       & 3s 3P*           & 3p 3D            & 1        & 3        \\
\noalign{\smallskip}
   5679.56 & N\,{\sc ii}                    &    5677.99 &      0.040 & $\pm  0.8$ &      0.031 & $\pm  1.6$ & V3       & 3s 3P*           & 3p 3D            & 5        & 7        \\
\noalign{\smallskip}
   5754.60 & [N\,{\sc ii}]                  &    5753.16 &      1.558 & $\pm  0.6$ &      1.167 & $\pm  1.5$ & F3       & 2p2 1D           & 2p2 1S           & 5        & 1        \\
\noalign{\smallskip}
   5801.51 & C\,{\sc iv}                    &    5799.40 &      0.081 & $\pm  1.6$ &      0.060 & $\pm  2.6$ & V1       & 3s 2S            & 3p 2P*           & 2        & 4        \\
\noalign{\smallskip}
   5812.14 & C\,{\sc iv}                    &    5814.25 &      0.148 & $\pm  1.4$ &      0.109 & $\pm  2.5$ & V1       & 3s 2S            & 3p 2P*           & 2        & 2        \\
\noalign{\smallskip}
   5875.66 & He\,{\sc i}                    &    5874.05 &     22.369 & $\pm  0.2$ &     16.262 & $\pm  1.4$ & V11      & 2p 3P*           & 3d 3D            & 9        & 15       \\
\noalign{\smallskip}
   5941.65 & N\,{\sc ii}                    &    5945.92 &      0.136 & $\pm  0.6$ &      0.097 & $\pm  1.7$ & V28      & 3p 3P            & 3d 3D*           & 5        & 7        \\
\noalign{\smallskip}
   5978.97 & S\,{\sc iii}                   &    5977.33 &      0.048 & $\pm  2.4$ &      0.034 & $_{ -3.0}^{+  3.4}$ & V4       &                  &                  &          &          \\
\noalign{\smallskip}
   6074.10 & He\,{\sc ii}                   &    6072.48 &      0.230 & $\pm  0.8$ &      0.159 & $\pm  1.9$ & 5.20     & 5g+ 2G           & 20h+ 2H*         & 50       & *        \\
\noalign{\smallskip}
   6101.83 & [K\,{\sc iv}]                  &    6099.88 &      0.017 & $\pm  3.7$ &      0.012 & $\pm  4.6$ & F1       & 3p4 3P           & 3d4 1D           & 5        & 5        \\
\noalign{\smallskip}
   6151.43 & C\,{\sc ii}                    &    6149.79 &      0.005 & $\pm  5.3$ &      0.003 & $\pm  6.5$ & V16.04   & 4d 2D            & 6f 2F*           & 10       & 14       \\
\noalign{\smallskip}
   6300.34 & [O\,{\sc i}]                   &    6298.70 &      6.098 & $\pm  0.3$ &      4.013 & $\pm  1.8$ & F1       & 2p4 3P           & 2p4 1D           & 5        & 5        \\
\noalign{\smallskip}
   6312.10 & [S\,{\sc iii}]                 &    6310.33 &      3.162 & $\pm  0.6$ &      2.075 & $\pm  1.9$ & F3       & 2p2 1D           & 2p2 1S           & 5        & 1        \\
\noalign{\smallskip}
   6363.78 & [O\,{\sc i}]                   &    6362.16 &      2.134 & $\pm  0.3$ &      1.385 & $\pm  1.8$ & F1       & 2p4 3P           & 2p4 1D           & 3        & 5        \\
\noalign{\smallskip}
   6371.38 & S\,{\sc iii}                   &    6369.56 &      0.096 & $\pm  1.3$ &      0.062 & $\pm  2.5$ & V2       & 4s 2S            & 4p 2P*           & 2        & 2        \\
\noalign{\smallskip}
   6548.10 & [N\,{\sc ii}]                  &    6546.44 &      8.179 & $\pm  0.8$ &      5.099 & $_{ -1.9}^{+  2.2}$ & F1       & 2p2 3P           & 2p2 1D           & 3        & 5        \\
\noalign{\smallskip}
   6583.50 & [N\,{\sc ii}]                  &    6581.80 &     24.956 & $\pm  0.7$ &     15.442 & $_{ -1.9}^{+  2.2}$ & F1       & 2p2 3P           & 2p2 1D           & 5        & 5        \\
\noalign{\smallskip}
   6678.16 & He\,{\sc i}                    &    6676.31 &      6.856 & $\pm  0.2$ &      4.159 & $_{ -1.8}^{+  2.0}$ & V46      & 2p 1P*           & 3d 1D            & 3        & 5        \\
\noalign{\smallskip}
   6716.44 & [S\,{\sc ii}]                  &    6715.18 &      1.124 & $\pm  2.0$ &      0.677 & $\pm  3.4$ & F2       & 2p3 4S*          & 2p3 2D*          & 4        & 6        \\
\noalign{\smallskip}
   6730.82 & [S\,{\sc ii}]                  &    6729.17 &      1.407 & $\pm  1.1$ &      0.844 & $\pm  2.6$ & F2       & 2p3 4S*          & 2p3 2D*          & 4        & 4        \\
\noalign{\smallskip}
   7065.25 & He\,{\sc i}                    &    7063.24 &     14.607 & $\pm  0.3$ &      8.214 & $\pm  2.3$ & V10      & 2p 3P*           & 3s 3S            & 9        & 3        \\
\noalign{\smallskip}
   7135.80 & [Ar\,{\sc iii}]                &    7133.89 &     13.598 & $\pm  0.5$ &      7.547 & $\pm  2.5$ & F1       & 3p4 3P           & 3p4 1D           & 5        & 5        \\
\noalign{\smallskip}
   7236.42 & C\,{\sc ii}                    &    7235.65 &      0.277 & $\pm  2.3$ &      0.151 & $_{ -3.7}^{+  4.2}$ & V3       & 3p 2P*           & 3d 2D            & 4        & 6        \\
\noalign{\smallskip}
   7237.17 & C\,{\sc ii}                    &    7235.22 &      0.113 & $\pm  4.2$ &      0.062 & $_{ -6.0}^{+  6.3}$ & V3       & 3p 2P*           & 3d 2D            & 4        & 4        \\
\noalign{\smallskip}
   7237.26 & [Ar\,{\sc iv}]                 &    7235.45 &      0.197 & $\pm  3.9$ &      0.107 & $\pm  5.9$ & F2       & 3p3 2D*          & 3p3 2P*          & 6        & 4        \\
\noalign{\smallskip}
   7254.38 & O\,{\sc i}                     &    7252.49 &      0.207 & $\pm  2.6$ &      0.112 & $\pm  4.5$ & V20      & 3p 3P            & 5s 3S*           & 3        & 3        \\
\noalign{\smallskip}
   7262.76 & [Ar\,{\sc iv}]                 &    7260.63 &      0.256 & $\pm  1.4$ &      0.139 & $\pm  3.3$ & F2       & 3p3 2D*          & 3p3 2P*          & 4        & 2        \\
\noalign{\smallskip}
\hline
\end{tabular}
\end{center}
\small
\begin{tablenotes}
\item[1]\textbf{Note.} Measured fluxes $F(\lambda)$ and dereddened fluxes $I(\lambda)$ with respect to H$\beta$ on a scale where H$\beta=100$.  The first column presents the laboratory wavelength $\lambda_{\rm lab}$, while the third column presents the observed wavelength $\lambda_{\rm obs}$.
\end{tablenotes}
\end{table*}

% Don't change these lines
\bsp	% typesetting comment
\label{lastpage}
\end{document}